\documentclass[12pt,preprint]{aastex}

\newcommand{\eg}           {{e.g.}}
\newcommand{\etal}         {{et~al. }}
\newcommand{\htwo}         {\hbox{H$_2$}}

\newcommand{\IRAS}         {\hbox{{\it IRAS\ }}}

\newcommand{\Spitzer}      {\hbox{{\it Spitzer\ }}}
\newcommand{\kms}          {\hbox{km$\,$s$^{-1}$}}
\newcommand{\approxlt}     {\lower.2em\hbox{$\buildrel < \over \sim$}}
\newcommand{\approxgt}     {\lower.2em\hbox{$\buildrel > \over \sim$}}
\newcommand{\Lco}          {\hbox{$L_{\rm CO}$}}
\newcommand{\Lhcn}         {\hbox{$L_{\rm HCN}$}}

\newcommand{\Lir}          {\hbox{$L_{\rm IR}$}}
\newcommand{\Lrc}          {\hbox{$L_{\rm RC}$}}
\newcommand{\ls}           {\hbox{L$_{\odot}$}}

\newcommand{\Msun}         {\hbox{M$_{\odot}$}}           

\begin{document}
\title{THE RADIO CONTINUUM, FAR-INFRARED EMISSION, AND DENSE 
MOLECULAR GAS IN GALAXIES}
\author{Fan Liu$^{1,2}$  and  Yu Gao$^1$}
\affil{1. Purple Mountain Observatory, Chinese Academy of
Sciences (CAS), 2 West Beijing Road, Nanjing 210008, P.R. China}
\email{yugao@pmo.ac.cn}
\affil{2. Graduate School of CAS, Beijing, 100080, P.R. China}

\begin{abstract}

 A tight linear correlation is established between the HCN line luminosity
 and the radio continuum (RC) luminosity for a sample of 65 galaxies (from Gao
 \& Solomon's HCN survey), including normal spiral galaxies and luminous and
 ultraluminous infrared galaxies (LIRGs/ULIRGs). After analyzing the various
 correlations among the global far-infrared (FIR), RC, CO, and HCN luminosities
 and their various ratios, we conclude that the FIR-RC and FIR-HCN
 correlations appear to be linear and are the tightest among all correlations.
 The combination of these two correlations could result in the tight RC-HCN
 correlation we observed.  Meanwhile, the non-linear RC-CO correlation shows 
 slightly larger scatter as compared with the RC-HCN correlation, and there is no
 correlation between ratios of either RC/HCN-CO/HCN or RC/FIR-CO/FIR. In 
 comparison, a meaningful correlation is still observed between ratios of 
 RC/CO-HCN/CO.  Nevertheless, the correlation between RC/FIR and HCN/FIR also
 disappears, reflecting again the two tightest FIR-RC and FIR-HCN correlations
 as well as suggesting that FIR seems to be the bridge that connects HCN with
 RC.   Interestingly,  despite obvious HCN-RC and RC-CO correlations, 
 multi-parameter fits hint that while both RC and HCN
 contribute significantly (with no contribution from CO) to FIR, yet RC is
 primarily determined from FIR with a very small contribution from CO and 
 essentially no contribution from HCN.  These
 analyses confirm independently the former conclusions that it is practical to
 use RC luminosity instead of FIR luminosity, at least globally, as an
 indicator of star formation rate in galaxies including LIRGs/ULIRGs,
 and HCN is a much better tracer of star-forming molecular gas and correlates
 with FIR much better than that of CO.

\end{abstract}

\keywords{galaxies: ISM --- galaxies: starburst --- infrared: galaxies
--- ISM: molecules --- radio continuum: galaxies --- radio lines: galaxies}

\section{INTRODUCTION}   
  
Ever since far-infrared (FIR) emission was thought to mainly originate as a
result of star formation in giant molecular clouds (GMCs; \eg, Mooney \& Solomon
1988), it had been used as a standard indicator of star formation rate
(SFR). But until now, the spatial resolution of FIR observation has limited
further detailed research, particularly in external galaxies. As another result
of the formation of young massive stars, though at the end of their rapid
formation and evolution, the radio continuum (RC) emission at centimeter (cm)
wavelength has a completely different emission mechanism compared with that of
the FIR emission.  RC offers a hope, with its spatially resolved capability, to
study SFR in galaxies both near (e.g., Condon et al. 1996) and far (e.g., 
Schinnerer et al. 2007) .

The FIR emission arises from the dust heated by new-born massive stars. The
young OB stars are imbedded in very tiny, yet massive and dense regions and all
of their UV/optical radiation is absorbed by dust which re-radiates mostly in
mid/FIR.  Though for normal disk galaxies the situation is more complex in that
the dust heating may also be contributed by older stellar population, and the
optical depth of dust may not be thick enough for the FIR luminosity to
truthfully measure the bolometric luminosity of entire galaxies except for
the nuclear starburst regions. However, the FIR emission should provide an
excellent measure of the SFR in luminous and ultraluminous infrared galaxies
(LIRGs/ULIRGs) where dusty circumnuclear starbursts dominate (Kennicutt 1998a).

The RC is mainly non-thermal synchrotron emission that arises from the
interaction of relativistic electrons with the ambient magnetic field in which
they diffuse. The supernova remnants (SNRs) of Type II and Type Ib supernovae
that are produced by massive stars (M$\gtrsim$8\Msun), which have lifetimes
$\lesssim 3 \times 10^{7}$~yr, are thought to have accelerated most of the
relativistic electrons. If we take the typical spiral disk field strength $B
\sim 5 \mu G$ and comparable magnetic energy density with inverse-Compton
radiation energy density, the synchrotron lifetime of the relativistic electrons
is $\lesssim 10^{8}$ yr while emitting at 1.5 GHz.  Therefore, RC can probe
very recent star formation activity in normal star-forming galaxies whose radio
emission is not dominated by the active galactic nucleus (AGN; Condon 1992).

On global scales, the FIR and RC emissions are linearly correlated over 5
orders of magnitude in luminosities for various galaxies from dwarf galaxies,
normal spiral galaxies, irregular galaxies, to starbursts, Seyferts, radio-quiet
quasars (Condon et al. 1991), and local LIRGs/ULIRGs (Yun, Reddy \& Condon 2001), out
to the most extreme star-forming galaxies in the early universe, at $z$ = 1 and
beyond (Appleton et al. 2004). The FIR-RC correlation for the \IRAS 2 Jy galaxy
sample (Yun et al. 2001, including 1809 galaxies) is well described by a linear
relation, and over 98\% of the sample galaxies follow this tight FIR-RC
correlation. This correlation offers a potential method of deriving the SFR
using the measured RC luminosity (Yun \etal 2001) even if the FIR luminosity is
unknown, and the RC images could potentially be used to judge how the FIR
distribution would look like on detailed small scales in galaxies (Condon \etal
1996). With
the high resolution capability of \Spitzer \ (and now {\it Herschel}), detailed
local FIR-RC correlation is also shown to be valid (\eg, Murphy \etal 2006, 2008).

Many researchers have conducted plenty of observations of CO emission in
galaxies, and the global correlations between FIR and CO (\eg, Devereux \& Young
1990; Young \& Scoville 1991), and between RC and CO (Rickard, Turner \&
Palmer 1977; Israel \& Rowan-Robinson 1984; Adler, Allen \& Lo 1991; Murgia
\etal 2002) are also well established for different samples. Adler \etal (1991)
and Murgia \etal (2002, 2005) found that the RC-CO correlation is linear from
global scale down to $\sim 100$ pc size scale, at which scale there is still no
evidence that this correlation is going to break down, with a dispersion that is
less than a factor of 2. Therefore, the mean star formation efficiency (SFE),
which measures the SFR (deduced from RC) per unit mass of molecular gas available to form
stars, is found to vary weakly with Hubble morphological type (among galaxies)
and distances from galaxy centers (within individual galaxy disks; Murgia \etal
2002).
    
For the HCN survey sample (Gao \& Solomon 2004a,b, hereafter GS04a,b) of
65 spiral galaxies, starbursts, LIRGs, and ULIRGs, we show here that this FIR and
RC correlation also holds and appears to be the best correlation among all correlations. 
Before the HCN survey of Gao \& Solomon (GS04b), the HCN emission in external
galaxies has only been observed in less than 30 galaxies, and only a few of them
are measured globally to derive the total HCN emission (\eg, Sorai \etal 2002;
Shibatsuka \etal 2003). Based on this systematic HCN survey, a tight linear
FIR-HCN correlation has been established (GS04a). Most recent follow-up
observations of dense molecular gas tracers, such as HCN, HCO$^+$, HNC, etc., in
galaxies (Baan et al. 2008; Gracia-Carpio et al. 2008) mostly confirm this
correlation. Some theoretical models also predict such linear correlations (\eg,
Krumholz \& Thompson 2007). According to this linear FIR-HCN correlation, a
Schmidt (1959) star formation law in terms of dense molecular gas is established
with a power-law index of 1.0. Furthermore, compared with the FIR-CO
correlation, the FIR-HCN is linear and tighter as the authors argued that the
combination of the stronger correlations between FIR-HCN and between HCN-CO may
account for the FIR-CO correlation. Based on these comparisons, they also argued
that the amount of the dense molecular gas traced by HCN, but not the total
amount of molecular gas traced by CO, could be a critical molecular parameter
that measures SFR in star-forming galaxies (GS04a). These new results from the
HCN survey obviously invoke a question on the comparison between the correlations
of RC-HCN versus RC-CO: is the RC-HCN correlation significantly better than the
RC-CO correlation?

How do the FIR and RC as SFR indicators relate with the star formation materials,
the molecular gas (CO), and particularly the dense molecular gas (HCN)? Stars are born mainly
in GMCs. The total mass of molecular gas of GMCs can be determined from the CO
luminosity. However, the excesses of the CO luminosity in GMC cores where active
high-mass star formation occurs are not specific enough to reveal the star
formation potential of the dense cores. The physical conditions of star-forming
GMC cores (Evans 1999) are better revealed by emission from
high dipole-moment molecules, like HCN, whose emission traces the dense molecular
gas ($n(H_2) \approxgt 3 \times 10^4$cm$^{-3}$) associated with the cores of the
star-forming GMCs (GS04b). FIR-HCN is shown to be linearly correlated (GS04a).
Does HCN also strongly correlate with RC?  What is the role of dense molecular
gas in the FIR-RC relationship?

Here, we utilize this HCN sample to analyze the various relationships among the
global HCN, FIR, CO, and RC luminosities to answer some questions raised here,
and to further demonstrate the possibility of using the RC luminosity instead of
FIR luminosity as an indicator of the SFR. The HCN sample and the total RC
luminosity of the sample are presented in Section 2. Section 3 presents the
results and analysis of the various correlations of these global luminosities
and their ratios and shows that the global correlation between RC and HCN
appears to be a combinational result of two tightest correlations between FIR-RC
and between FIR-HCN even though RC-HCN correlation is significantly better than
the RC-CO correlation. Discussion on the possible physical relationship between
the HCN (dense molecular gas tracer) and the RC emission (the indicator of the
rate of high-mass star formation) in galaxies is presented in Section 4,
followed by the main points of this study in Section 5.

\section{THE RADIO CONTINUUM SAMPLE}
   
   Our RC sample is drawn entirely from
the HCN sample of Gao \& Solomon (GS04a,b). The sample
includes almost all galaxies with strong CO and FIR emission
in the northern sky (53 in total, GS04b) and 12 additional
galaxies (mostly LIRGs/ULIRGs) from literatures, mainly from
Solomon, Downes \& Radford (1992). Details about the sample
properties and the measurements of the total HCN emission in
galaxies can be found in GS04b.

The RC flux is mainly obtained from the NRAO VLA Sky Survey (NVSS; Condon et
al. 1998), which covers the sky north of J2000.0 decl.=$-40\arcdeg$ at 1.4 GHz
(Condon et al. 1998), whereas for some large sources with diameter $D_{25}$
larger than $\sim 8\arcmin$, we adopted the fluxes given in previous RC surveys
(e.g., Segalovitz 1977; Condon 1987; Condon \& Broderic 1988; Condon et
al. 1990, 1996) to avoid the missing flux problem that NVSS suffers for sources
with diameters larger than $\sim 8\arcmin$.  The flux of NGC~4945, which is a
southern galaxy not covered by NVSS, is taken from the Parkes catalogue as
listed in the NED\footnote{NASA/IPAC Extragalactic Database(NED) is operated by
  the Jet Propulsion Laboratory, Caltech, under contract with the National
  Aeronautics and Space Administration.}.

   The derived global luminosity of RC at 1.4 GHz and
HCN, FIR, and CO (from GS04a,b) luminosities of galaxies in this
sample are listed in Table 1 (here we make no distinction between
the total IR and FIR emission, though the total IR luminosity defined
in Sanders \& Mirabel 1996 is used). A few sensitive HCN limits are
kept in Table 1 and included for various analyses.

   The largest uncertainties are of the HCN observations, as
described in detail in GS04b, at a level of up to
$\sim$30\%, whereas for several large nearby sources without
sufficient mapping the uncertainties can be as large as
$\sim$50\% ($\sim$20\% calibration error included). The
uncertainties of CO luminosities are mainly dominated by the
calibration errors and of level $\sim$20\%. FIR
luminosities from \IRAS and RC luminosities from NVSS are
much more accurate; the uncertainties are of a few percent
only.

   For several large sources in the HCN survey sample, e.g.,
NGC253, only the nuclear regions and the innermost disks are
mapped in HCN , since they are limited by the sensitivity of the
instruments, while a rough estimate of the NVSS image shows
that $\gtrsim 70\%$ RC emission is from the same region.
Obviously, tighter constraints in various correlations can
be obtained once more sensitive and accurate HCN
observations are conducted with more powerful telescopes,
such as GBT\footnote{The Robert C. Byrd Green Bank Telescope
(GBT). http://www.gb.nrao.edu/gbt/}\&LMT\footnote{The Large
Millimeter Telescope (LMT). http://www.lmtgtm.org/}, both of which are
soon to be available.

\section{RESULTS AND ANALYSES}

\subsection{ RC--HCN Correlation}
   
   Similar to the FIR-HCN correlation (GS04a), the principal
result from this research is the tight linear correlation
between RC luminosity \Lrc~ and HCN line luminosity \Lhcn~
(Figure 1(a)). The correlation extends over 3 orders of
magnitude in luminosities and includes normal spiral
galaxies, starbursts, and LIRGs/ULIRGs. Note that there appears
to be no systematic difference between host galaxies with
known AGN and the rest of the galaxies in the RC-HCN correlations.
A linear regression yields a power-law slope of 0.99($\pm$
0.05). The corresponding correlation coefficient is $R = 0.92$
($R^2 = 0.84$) for \Lrc--\Lhcn. The best-fit (logarithmic)
relation between HCN and RC luminosities is

\begin{equation}
log \Lrc = 0.99(\pm 0.05) log \Lhcn - 2.0.
\end{equation}
   
   Compared with the FIR-HCN result  (GS04a), 
\Lrc-\Lhcn \  is an almost equally tight linear correlation as that
of the \Lir-\Lhcn.
   
   When we use all galaxies excluding 6 HCN limits, the
correlation remains almost the same with $log \Lrc =
0.98(\pm0.07) log \Lhcn -1.9$, and essentially the same
correlation coefficient $R = 0.90$ ($R^2 = 0.81$).
   
   The RC-CO correlation is also very strong ($R = 0.89$, $R^2
= 0.79$, shown in Figure 1(b)) and appears to be nearly as
tight as the RC-HCN correlation. But similar to the FIR-CO
correlation, the RC-CO correlation is nonlinear. It is also noticeable 
that the range of CO luminosity is smaller covering only a bit over
2 orders of magnitude, whereas both HCN and RC luminosities 
spread over 3 orders of magnitude. The
comparison of the luminosity ratios of \Lrc/\Lco \ and
\Lrc/\Lhcn \ with \Lrc~ \ shows a dramatic difference: the
\Lrc/\Lco \  ratio increases substantially with increasing
\Lrc, whereas \Lrc/\Lhcn \ appears to be nearly independent
of \Lrc~ (Figures 2(a) and (b)). These are nearly the same plots
with \Lir \ instead of \Lrc \ shown in GS04a,
demonstrating the tight linearity of the correlation between \Lrc \ and
\Lhcn \ in contrast  to the  nonlinearity of the correlation between \Lrc \ and \Lco \ as we further
illustrate below.

\subsection{Comparison of Correlations Between the Ratios}   

Figure 3(a) shows a correlation between luminosity ratios \Lrc/\Lco \ and
\Lhcn/\Lco \ with a correlation coefficient $R = 0.63$ ($R^2 = 0.40$). This
suggests that the new star formation efficiency (SFE=\Lrc/\Lco, i.e., SFR --- derived 
by using the RC luminosity --- per unit of molecular gas mass inferred by CO) 
depends on the fraction of molecular gas in a
dense phase (indicated by \Lhcn/\Lco \ ). This correlation better demonstrates
the connection between the RC and HCN luminosities (as shown in Figure~1). 
Both \Lrc \ and \Lhcn \ have been normalized by \Lco \ to better show the
possible physical relationship between \Lrc \ and \Lhcn \ after removing the
dependence upon the distance, galaxy size, and other possible selection
effects.
   
Similarly, we can show the correlation between \Lrc \ and \Lco \ divided by
\Lhcn \ for normalization (Figure 3(b)).  Same as the correlation between \Lir
\ and \Lco \ normalized by \Lhcn, the apparently strong correlation between
\Lrc \ and \Lco \ (Figure 1(b)) also disappeared once normalized by \Lhcn~ 
($R = 0.20$, $R^2 = 0.04$).  This reflects the tight linearity of the correlation between RC and HCN
(Figure 1(a)), making the ratio RC/HCN nearly constant. 

These are essentially the same arguments and procedures as in GS04a in comparing the
FIR-HCN versus FIR-CO that shows a fundamental difference between HCN and CO
(star-forming dense molecular gas and total molecular gas) in their relationship 
to FIR (SFR). Therefore, the difference between HCN and CO in their relationship 
to SFR is again clearly shown here using RC instead of FIR as a proxy of SFR. 
Meanwhile, the correlation between \Lrc \ and \Lir \ after normalization by \Lhcn
\ (Figure 4(a)) remains rather tight ($R = 0.74$, $R^2 = 0.55$) though there are
very tight RC-HCN and FIR-HCN correlations, which
suggests that the FIR-RC correlation is indeed the tightest among all correlations.
   
Nevertheless, the tight correlation between \Lrc \ and \Lhcn \ after
normalization by \Lir \ (Figure 4(b)) has almost completely disappeared, too
(c.f., Figure~3), whereas the correlation between \Lir \ and \Lhcn \ divided by
\Lrc \ for normalization (Figure 4c) still remains as a meaningful
correlation ($R = 0.57$, $R^2 = 0.33$, and even a stronger correlation if they
are normalized by CO instead of RC as shown in GS04a). This certainly shows some subtle
differences in the tight correlations among FIR, RC, and HCN luminosities and
surely implies that the FIR-HCN correlation is much better than the RC-HCN 
correlation. As further demonstrated with multi-parameter fits and discussed in
Section~4, FIR and HCN might have a better direct physical relationship with SFR than
other quantities involved in this paper.  The RC-HCN correlation is likely a result
of the combination of the two tightest 
FIR-RC and FIR-HCN correlations. This point can be further
revealed and corroborated by comparing these correlations with the
FIR/RC-CO/RC (FIR-CO normalized by RC) and FIR/CO-RC/CO (FIR-RC normalized 
by CO) correlations shown in Figure~5.
   
Using the RC luminosity as the normalization, the correlations between the FIR
luminosity and HCN (Figure~4(c)) and CO (Figure~5(a)) luminosities still show weak, but
meaningful correlations with the correlation coefficient $R =$ 0.57, 0.37 ($R^2
= 0.32, 0.14$), respectively, though the latter FIR/RC-CO/RC correlation 
is very marginal.  Normalized by HCN (Figure~4(a)) and CO (Figure~5(b))
luminosities, the RC and FIR luminosities still show the tightest correlations
with the correlation coefficient $R = 0.74$, 0.85 ($R^2 = 0.55, 0.72$),
respectively. Normalized by the FIR luminosity, however, the RC luminosity
does not show any correlation with CO (not shown) or with HCN (Figure~4(b))
luminosities ($R = 0.17$, 0.14, $R^2 = 0.03, 0.02$, respectively). Although both 
luminosity ratio correlations are still quite tight, noticeable difference is obviously 
shown between Figure~3(a) (or FIR-HCN normalized by CO; Figure~5(a) in GS04a) and 
Figure~5(b) that the FIR/CO-RC/CO correlation appears to be the tightest
among all luminosity ratio correlations. 

In summary, all correlations shown above suggest that only FIR-RC and FIR-HCN
correlations are the tightest ones.  The RC-HCN correlation is much tighter
than the RC-CO correlation, yet the former one can be expected from the strongest
FIR-RC and FIR-HCN correlations.  We will further demonstrate these results by the three-
and four-parameter fits in Section 3.3.  For the sake of completeness,
we include the result of the well known FIR-RC correlation and other directly related 
correlations for this HCN sample in Appendix A.

\subsection{Multi-parameter Fits}

Following the practices of the model parameter fits in GS04a, which
have demonstrated that HCN is a much better active star formation tracer than
CO, we here similarly discuss the three- and four-parameter fits,
respectively, involving the RC, FIR, CO, and HCN luminosities.

\subsubsection{The Three-parameter (FIR, HCN, and RC) Fits}

Given that the tightest correlations are among FIR, HCN, and RC, we discuss here these 
three-parameter fits first, including FIR, RC, and HCN luminosities. We also list other
three-parameter fits in Appendix B for comparison and completeness.

   The RC luminosity from a model fit to the HCN and FIR (the
RC(\Lhcn, \Lir \ ) model) yields

\begin{equation}
   log \Lrc(\Lhcn, \Lir) = (0.10 \pm 0.11) log \Lhcn + (0.90 \pm 0.10) log \Lir - 4.80.
\end{equation}

The extremely weak dependence (almost independent) on \Lhcn \ shows that the RC
luminosity is determined principally from the FIR luminosity with only extremely
marginal contribution from HCN. This seems to be a bit odd given the tight RC-HCN
correlation shown in Equation (1). Nevertheless, this does imply that HCN contributes
insignificantly and is almost random once FIR is fitted for RC.
   
An RC and FIR luminosity model for \Lhcn \ (the HCN(\Lir, \Lrc) model) gives

\begin{equation}
   log \Lhcn(\Lir, \Lrc) = (0.14 \pm 0.15) \Lrc + (0.77 \pm 0.15) \Lir - 1.20.
\end{equation}

The \Lhcn \ luminosity is mainly determined from the FIR luminosity and is only
weakly dependent on the RC luminosity. This model fit basically shows that RC contributes
little to HCN  after fitting FIR for HCN, despite the strong correlation between RC and HCN.
   
The FIR luminosity from a model fit to the HCN and RC (the IR(\Lhcn, \Lrc)
model), however, yields

\begin{equation}
   log \Lir(\Lhcn, \Lrc) = (0.38 \pm 0.08) log \Lhcn + (0.61 \pm 0.07) log \Lrc + 4.30.
\end{equation}

Here, \Lhcn \ and \Lrc \ seem to be roughly comparably important in predicting FIR, 
though RC appears to have more weight than HCN.  We caution here that the usually more 
than 10 times larger errors in HCN than that in RC could easily be a cause of this
slight disfavor for HCN. This relation produces a much tighter fit
than any of the simple two-parameter fits, including the tightest FIR-HCN and FIR-RC
correlations.

\subsubsection{The Four-parameter Fits}

CO usually contributes the least whenever HCN is involved in the three-parameter 
(FIR, HCN, CO) fits (GS04a). Will this still be true if an additional parameter, RC, is added?
We here discuss the four-parameter fits.

The RC luminosity from a model fit to the HCN, FIR, and CO (the RC(\Lhcn, \Lir,
\Lco \ ) model) yields

\begin{equation}
   log \Lrc(\Lhcn, \Lir, \Lco) = (-0.02 \pm 0.13) log \Lhcn + (0.87 \pm 0.10)
   log \Lir + (0.22 \pm 0.13) log \Lco - 5.55.
\end{equation}

The extremely weak dependence (almost independent) on \Lhcn \ shows that the RC
luminosity is determined principally from the FIR luminosity with only marginal
contribution from the CO and essentially nothing from the HCN.
   
An RC, CO, and FIR luminosity model for \Lhcn \ (the HCN(\Lir, \Lrc, \Lco)
model) gives

\begin{equation}
  log \Lhcn(\Lir, \Lrc, \Lco) = (-0.02 \pm 0.13) log \Lrc + (0.58 \pm 0.14) log \Lir +
  (0.55 \pm 0.11) \Lco - 3.37.
\end{equation}

The \Lhcn \ luminosity is independent($\sim -0.02$) of RC luminosity, and
is almost equally determined by FIR and CO luminosity.
   
The FIR luminosity from a model fit to the HCN, RC, and CO (the IR(\Lhcn,
  \Lrc, \Lco) model), however, yields

\begin{equation}
   log \Lir(\Lhcn, \Lrc, \Lco) = (0.40 \pm 0.09) log \Lhcn + (0.62 \pm 0.07) log
   \Lrc - (0.04 \pm 0.11) log \Lco + 4.49.
\end{equation}

Here, \Lhcn \ and \Lrc \ seem to be roughly comparably important though RC
appears to have more weight than HCN. This is much more extreme than the 
three-parameter fit (HCN+CO for FIR) given in Equation (3) of GS04a, and CO 
indeed contributes nothing to FIR when both HCN and RC are used in the 
fit.  In other words, adding CO does not have any effect on predicting FIR once
both HCN and RC are involved.

Finally, we show the CO luminosity from a model fit of the HCN, RC, and FIR
(the CO(\Lhcn, \Lrc, \Lir) model), which yields

\begin{equation}
   log \Lco(\Lhcn, \Lrc, \Lir) = (0.51 \pm 0.10) log \Lhcn + (0.21 \pm 0.12) log
   \Lrc - (0.06 \pm 0.14) log \Lir + 4.54.
\end{equation}

The \Lco \ luminosity is independent ($-0.06$) of FIR luminosity, and is
mainly determined by \Lhcn (0.51) and marginally by \Lrc \ 
(0.21). This is essentially very similar to Equation (5) in GS04a, i.e., CO 
is mainly determined by HCN with  little contribution from FIR.
Adding an additional parameter RC only affects the overall fit slightly and 
makes FIR useless in predicting CO from HCN.
Although it is marginal, this might 
suggest that RC is slightly more important than FIR in predicting CO.

In summary, FIR-HCN and FIR-RC correlations are the strongest ones among all. 
Despite a tight RC-CO correlation, the RC-HCN correlation is much tighter in
comparison. Nonetheless, it appears that the RC-HCN correlation might be 
a direct consequence of the combinational result from the two tightest RC-FIR 
and FIR-HCN correlations. The multi-parameter fit models further support these 
results and reveal the difference between HCN and CO in predicting FIR and 
RC. Being the tightest correlations for both FIR-RC and FIR-HCN, these 
results corroborate that it is practical to use RC instead of FIR to indicate the global
SFR, and HCN is much more important than CO in relating to FIR.

\section{DISCUSSION}

\subsection{The Most Fundamental Correlation: FIR-RC versus FIR-HCN}

   In the RC-HCN correlation shown in Figure~1(a), no significant differences can be
seen between the AGN host galaxies and others (the galaxies with AGN embedded are
represented by stars, a slight excess in RC appears to be present at high 
luminosity end). This indicates that these AGNs (except those in ULIRGs) do not show 
stronger excess in RC than the rest of the star-forming galaxies. Namely, AGN
contribution to RC is, on average, presumably roughly less
than half of the total RC emission, and not much
systematic difference that can be seen globally in this HCN sample
of star-forming galaxies and LIRGs/ULIRGs (GS04a). Similarly,
there is also no prominent difference between the AGN host galaxies and 
the rest of the star-forming galaxies in the FIR-HCN correlation (GS04a).

   The correlation between global FIR and RC luminosities is
found to be tight and linearly valid over 5 orders of
magnitude in thousands of galaxies (\eg, Yun \etal 2001).
This is also reinforced here by the HCN survey sample even though
the ranges in luminosities are over only 3 orders of
magnitude due to very limited sample size. Moreover, radio AGNs
usually show distinct excess in RC over FIR as compared to the general populations
of normal galaxies in the FIR-RC correlation. But, AGNs in the HCN sample do not 
particularly show any significant radio emission excess. As detailed above, it turns out that
the FIR-RC is the tightest correlation among all. But the FIR-HCN correlation 
is essentially as tight as the FIR-RC correlation, and these two correlations 
are the tightest. It might even be possible that the FIR-HCN correlation could be 
tighter than the FIR-RC correlation, especially in view of the significantly 
larger uncertainties in the estimate of HCN than other parameters.  This is also
intuitively implied since both FIR and HCN are directly and physically
related to active massive star formation. 

Several models were previously posed to interpret this correlation (\eg,
Voelk 1989; Helou \& Bicay 1993; Niklas \& Beck 1997; Murgia
\etal 2005). Most recently, Lacki, Thompson, \& Quataert (2009) suggest that
the FIR-RC correlation is a combinational result of the efficient cooling of 
cosmic-ray (CR) electrons in starbursts and a conspiracy of 
several factors: the decrease in the radio emission due to bremsstrahlung, ionization, and 
inverse Compton cooling in starbursts is countered by secondary electrons/positrons and the decreasing critical synchrotron frequency, which both increase the radio emission; 
 low effective UV dust opacity leads to the decreasing FIR emission, which balances the 
 decrease in radio emission caused by CR escape for lower surface density galaxies.
In fact, this kind of calorimetry  theory that
galaxies act in a reasonable approximation as calorimeters for the stellar UV radiation 
and for the energy flux of the CR electrons was proposed 20 years ago (Voelk 1989).

Although the molecular gas connection to the FIR-RC
correlation in galaxies is proposed by Murgia \etal (2005),
we demonstrated here that the addition of the total dense molecular
gas traced by HCN appears
to be globally insignificant in providing the dense star-forming gas connection between
the FIR (SFR) and RC (the RC(\Lhcn, \Lir, \Lco \ ) model, Equations (2), (5), and B(3)).  
Yet, the contribution of CO is non-negligible and appears to be more important than HCN.
Nevertheless, the star formation materials, particularly the dense gas (HCN), 
definitely play an important role in the obvious connection of SFR
to both the FIR and RC emission. Both HCN and RC contribute nearly 
equally to FIR and there is no additional contribution to FIR by adding 
CO (the IR(\Lhcn,  \Lrc, \Lco) model, eqs. (4), (7) and B(3)).  Thus, CO seems to be the least 
important one among all the four parameters RC, FIR, CO, and HCN
in relating to SFR even though they all show quite good correlations
with each other.

\subsubsection{The Origin of the Tightest Correlation: FIR-RC and FIR-HCN}

   In conventional models, the RC emission produced from
synchrotron radiation is related with massive star formation
via several steps: star formation $\rightarrow$supernovae
and SNRs $\rightarrow$relativistic
electrons accelerated by SNRs.  So, too,  is the FIR emission related: star
formation$\rightarrow$UV radiation$\rightarrow$absorption
and re-emission by dust enshrouding new-born stars. These should
all be expected to be the dominant  FIR and RC emission in active 
star-forming galaxies. Nevertheless, CR electrons involve with other
radiation processes, generating mostly the background RC emission that is
not related to massive star formation (the general interstellar radiation field). 
Additionally, the old stars also heat up the dust and additionally contribute
to the general interstellar radiation field.
The contribution of the general interstellar radiation field to 
the total FIR emission in galaxies might be significant at the 
low-\Lir\ end, where the general infrared interstellar radiation 
field is comparable to or even more dominant than the FIR radiation from the active 
star formation.

The temporal and spatial inconsistencies between these processes
could be the obstacles to understand fully the physical
origin of the FIR and RC correlation. Given the obvious star
formation connection of both FIR and RC and the important
role of the star formation materials, i.e., the molecular gas,
especially the dense molecular gas, in actually giving births of
massive stars, it can surely help us figure out which of these
correlations are more fundamental, and which are possibly
indirectly less fundamental correlations. Globally, we have seen the bigger role 
of HCN than CO in the FIR-RC correlation. Locally, the spatially resolved studies 
perhaps help us much better to make the whole picture
clearer, by taking the molecular gas and dense molecular
gas into account in the analysis of various comparisons of
the correlations.

 Temporally, massive stars live $\sim 10^{7}$ yr and the
 relativistic electrons probably have lifetimes of $\sim 10^{8}$ yr in
 star-forming galaxies; the RC luminosity therefore probes star formation
 activity not much earlier than $10^{8}$ yr ago. Taking the molecular gas in
 the dense phase as the initial stage in initiating the massive star formation
 procedure, and the total molecular gas as the potential star formation gas
 reservoir as a necessity in eventually forming the dense cores of GMCs, we
 can sort the three probes of SFR into a time line: CO (GMCs, assembly of star
 formation material---a necessity for later core collapse and star formation in
 GMCs)$\rightarrow$HCN (probes dense GMC cores, massive star formation
 sites)$\rightarrow$FIR (probes new-born young massive stars)$\rightarrow$RC
 (probes the end products of the short-lived massive stars).  If the dense
 cores of GMCs are forming massive stars on a time-scale smaller than the
 depletion time of dense gas in galaxies ($\leq 10^7$ yr; GS04a), followed by the life span of massive
 stars, then this is phenomenologically consistent with our result that only
 HCN-FIR and FIR-RC correlations are the tightest and have the least scatter
 from linearity, while HCN and RC luminosity appears to be the result of a
 combination of the former two correlations, not an immediate relation in time. 
 The obvious contrast between Figure~3(a) (or FIR-HCN normalized by CO (Figure~5(a) in GS04a)) 
 and Figure~5(b) might be reflected in the differences in these time-scales that it
 is much more dynamic and scattered in luminosity ratios involving HCN rather than CO.

Spatially, CO traces large-scale molecular gas distribution (entire GMCs plus
diffuse molecular clouds), where most of the molecular gas is not forming stars,
whereas the high-density regions (dense GMC cores traced by high-dipole moment
molecules such as HCN) of much smaller spatial scales are indeed the locations
of active star formation in galaxies. Murgia \etal (2005) showed that the
spatially resolved tight CO-RC correlation holds down to $\sim 100$pc size scale in galaxies
which is close to the GMC size scale. Paladino et al. (2006) further probed
the RC-FIR-CO correlations down to linear scales of a few hundred pc using new
{\it Spitzer} IR images of six BIMA CO Survey of Nearby Galaxies (BIMA SONG, Helfer et
al. 2003) and observed local deviations from the correlations in regions with
a high SFR where a low RC/FIR ratio is found. We showed here, however, that
the global HCN-RC correlation is actually much tighter than the CO-RC
correlation, but we need a similarly detailed, spatially resolved local
comparison between RC and HCN, and FIR and HCN in galaxies in order to examine
how the FIR-RC-HCN correlations extend to much smaller size scale.

  Murphy \etal (2006, 2008) have taken an initial look at the FIR-RC
  correlation within the disks of 4, and later increased to 29, nearby
  face-on galaxies in the {\it Spitzer} SINGS legacy program (Kennicutt
  \etal 2003), and found the trend that the ratio of FIR to RC decreases with
  increasing radius, which is consistent with what Marsh \& Helou (1995) found
  at intermediate spatial resolution.  They also studied how the star formation
  activity affects the FIR-RC correlation within galaxies by testing a
  phenomenological model which smears the FIR images to match the radio images.
  They found that the mean distance traveled by the CR electrons is most
  sensitive to the dominant age of the CR electron population, rather
  than the interstellar medium (ISM) parameters, which may inhibit their propagation, such as the ISM
  density, radiation-field energy density, and magnetic field strength.
  Comparison of such detailed spatially resolved correlations in line with our findings 
  in the global quantities could help us reach our final goal, i.e.,  understanding the FIR-RC
  correlation. Globally, HCN seems better than CO,  the validity of correlations of RC-CO and 
  FIR-RC on small scale has already been proven; we need high-fidelity HCN imaging and/or 
  resolved HCN observations to truly compare all.

Both FIR and RC emission involves physical processes of both small and large
spatial scales even though most of the FIR emission is dominated by active
star-forming regions of small scales. Mooney \& Solomon (1988) showed that
the FIR-CO correlation for GMCs improves, once the diffuse FIR emission that originated
from the general interstellar radiation field of large spatial scale was
subtracted. RC might be dominated by large-scale shocks/bubbles associated
with SNRs as well as even larger scale of magnetic fields, 
where CR electrons pass along the field lines and experience efficient cooling, 
whereas HCN traces a smaller
size scale than that of FIR and RC. Yet, they are associated with three
different periods in the time sequence connected with the entire star
formation processes. FIR emission originates from the dust that enshrouds the
new-born stars while HCN emission outlines regions of dense molecular gas that
eventually nurse new-born stars, whereas RC emission is produced from the
significantly diffused CR electrons and large-scale shocks, which have
traveled a long distance from the previous star-forming sites. The difference
between these three emissions in time sequence is in de facto agreement
with that of the corresponding locations in spatial scales: dense molecular
cores further collapse to form massive stars that then quickly evolve and go
through supernovae to become SNRs.

   In short, CO and FIR emission are almost entirely
associated with the GMCs, but most FIR emission is probably
associated with star-forming sites inside the dense cores of
the much smaller scales traced by the HCN emission. On the
other hand, RC is probably dominated by large-scale diffuse
emission though some FIR emission is also associated with the
large-scale general interstellar radiation field. These
differences are possibly suggested by the multi-parameter fits
that RC correlates much tighter with FIR than with HCN though CO
seems more important than HCN in predicting RC (Equations (2) and (5)),
and HCN correlates much tighter with FIR than with RC, though 
CO seems to be as equally important as FIR in predicting HCN (Equations (3) and (6)). 
Both FIR and HCN trace smaller scale star-forming regions, whereas RC
traces the overall large-scale environment, which are affected
by the feedback of massive star formation and previously
hosted massive stars.

\subsection{Radio Continuum (RC, and FIR and HCN) as Star Formation Rate (SFR) Indicator}

   Gao \& Solomon (GS04a) have extensively discussed
the dense molecular gas in relationship to the total
molecular gas and SFR indicated by the FIR emission. The HCN
emission is associated with the high-density molecular gas
which is the direct active star formation material. Compared with CO, 
the HCN luminosity is much better at
predicting the FIR luminosity for all galaxies including
ULIRGs (GS04a). This is represented by a much better
\Lir--\Lhcn \ correlation than  the \Lir--\Lco \ correlation
(GS04a). A similar result is obtained here by the comparison
between the \Lrc--\Lhcn \ correlation and the \Lrc--\Lco \
correlation. Although the extreme claim that globally, the RC-CO correlation
is as good as the FIR-RC correlation are made by Murgia \etal
(2005), perhaps due to a small range of parameters and limited
sample size, we find here that RC luminosity correlates with
\Lhcn \ much more tightly than with \Lco, and that the FIR-RC and
FIR-HCN are the tightest correlations among all. The physical
explanation for the tight correlation between the HCN and FIR
is obviously that stars are formed in dense molecular gas,
whereas the tight correlation between HCN and RC would
further strongly support this interpretation and at least
globally indicate that the RC could be used as a tracer of
star formation. For a detailed discussion on SFR, dense
molecular gas, total molecular gas, and their various
ratios readers are referred to Section 4 in Gao \& Solomon
(GS04a).

   According to the tight FIR-RC correlation which is also
valid down to kpc/sub-kpc size scale within galaxies (e.g.,
Marsh \& Helou 1995; Lu et al. 1996; Murphy \etal 2006, 2008),
detailed SFE maps can be deduced by using local RC
luminosity as a local SFR tracer and comparing them to the CO
images. This application has been first used in studies of
individual interacting galaxies, like Arp 244---pioneered by 
Gao \etal (2001)--- and Taffy galaxy (Gao, Zhu, \& Sequist 2003;
Zhu \etal 2007). These SFE maps allow us to identify
the most active star-forming sites, and characterize and investigate the star
formation properties in local regions of kpc/sub-kpc size
scales.

   Nonetheless, the mean SFE, which measures the SFR
(deduced from RC) per unit mass of molecular gas available
to form stars, is found to vary weakly with Hubble
morphological type (among galaxies) and distances from
galaxy centers (within individual galaxy disks; Murgia
\etal 2002). Murgia \etal (2005), however, do not measure any
systematic trend in the CO/RC ratio as a function of radius in the 
nine BIMA SONG galaxies studied. A hydrostatic pressure
regulation model was used to interpret the excellent
correlation between the CO, RC, and FIR emissions in galaxies
on both large and small scales to avoid invoking any
explicit dependence on the star formation scenario (Murgia \etal
2005). Given the assumption that CO surface brightness is
proportional to molecular gas surface density, the model
predicts $I_{\rm RC} \propto I_{\rm CO}^{1.4}$, which is consistent
with our global result and also, probably results from the
linear correlation between FIR-RC, consistent with the 1.4
power in global FIR versus CO (GS04a) and FIR versus (\htwo +HI) (Kennicutt
1998b) as well as recent local resolved studies (Calzetti et al. 2007; Kennicutt et al. 2007). 
We note that if the observational result that
$\Lhcn = 1.38 log \Lco -4.79$ (GS04a) is used, the prediction of this
model will become $I_{\rm RC} \propto I_{\rm HCN}$, also perfectly
consistent with the result of our global correlation fits.

   However, Paladino \etal (2006) observed local deviations from
the RC-IR-CO correlations in regions with a high SFR, such
as the spiral arms, in six galaxies for which
high-resolution {\it Spitzer} 24$\mu$m mid-IR data are available. Further
studies in a variety of star-forming galaxies are necessary
to enlarge the dynamical range of parameter spaces and to
achieve better agreements in the local RC-FIR-CO and even
RC-FIR-CO-HCN correlations. Nevertheless, Paladino \etal (2006) 
concluded that, down to $\sim 100$ pc scale, they have not yet probed the 
physical scales at which the correlations break down. 
 
   Based on \Spitzer mid-IR observations, Wu et al. (2005)
find that the 8$\micron$ and 24$\micron$ luminosities of
star-forming galaxies are both strongly correlated with
their RC (1.4GHz) and H$\alpha$ luminosities over a range in
luminosities of nearly 3 orders of magnitude. This
suggests that, alternatively, mid-IR emission of much better
spatial resolution than FIR can be approximately used as local tracers
of SFR as well. Yet, it is still unclear how quantitatively different
the various correlations between each SFR indicator: from mid-IR to FIR, and 
from RC to molecular gas tracers CO and HCN, could be, on local sub-kpc scales. 

With the help of the high-resolution IR capability of \Spitzer, and
now {\it Herschel}, as well as high-resolution and high-sensitivity HCN and 
CO imaging (with the upcoming ALMA), we can
further analyze these correlations down to sub-kpc scales in galaxies and
extremely smaller scale (sub-pc) inside the dense GMC cores in the Milky Way.
Whether these relations hold or break down will help make our understanding
of star formation in both external galaxies approaching microscopic scales and
the active star-forming regions in our Galaxy more comprehensive.

\section{SUMMARY}
   
   We summarize our main results and present our concluding
remarks in the following.

1. For the HCN survey sample of 65 galaxies, including normal galaxies and
LIRGs/ULIRGs, the most
luminous galaxies in the local universe, a tight linear correlation between the
global RC and HCN luminosities is established. This 
correlation is comparably tight as
that between FIR and HCN and is valid over 3 orders of magnitude including
ULIRGs. Normalized by \Lco, the correlation between RC and HCN is still
prominent, which suggests that the SFE (\Lrc/\Lco) is clearly proportional to
the dense molecular gas fraction (\Lhcn/\Lco). Compared with the same trend
found between the FIR-derived SFE and dense molecular gas fraction, this
consistency supports these suggestions from another point of view.

2. We also observe the tight correlation between RC
and total FIR luminosities, which is well established in previous studies of
much larger samples.  This FIR-RC correlation is the tightest one (with a
correlation coefficient $R = 0.96$) among the various correlations between RC,
FIR, CO, and HCN luminosities discussed in this paper. Moreover, there is also
a significant correlation between the normalized luminosities \Lrc/\Lhcn \ and
\Lir/\Lhcn \ even though FIR is almost equally tightly correlated with HCN.  This confirms
the well established FIR-RC correlation independently with our HCN sample
though the detailed physical processes that relate them are still not fully
understood.

3. The three- and four-parameter fits show that RC correlates much tighter
with FIR than with HCN and that HCN correlates much tighter with FIR than with
RC. Therefore, the combination of the well-known strongest FIR-RC correlation
and perhaps the tightest linear correlation between FIR and HCN luminosities accounts
for the RC-HCN correlation. With the support of other various FIR-RC-HCN-CO
correlations discussed in this paper, we corroborate that at least on global
scale it is practical to use RC luminosity instead of FIR luminosity as an
indicator of SFR, and HCN is much more important than CO in predicting to FIR.
  
4. The various correlations are discussed under the current understanding of
the star formation process, from raw material for star formation to star birth
and death. Our global correlation results show that HCN relates tighter with
SFR tracers FIR and RC as compared with CO, and that FIR seems to be the
bridge that connects HCN with RC.  These correlations are roughly consistent with the 
physical picture of massive star formation processes spatially and temporally.

\acknowledgments
  
   We thank the anonymous referees for helpful comments, which
   improved the presentation of this paper. This research has made use of the
   NRAO VLA SKY SURVEY database and NASA/IPAC Extragalactic Database
   (NED). Research for this project is partly supported by NSF of China
   (Distinguished Young Scholars \#10425313, grants \#10833006 and \#10621303),
   Chinese Academy of Sciences' Hundred Talent Program, and 973 project of the
   Ministry of Science and Technology of China (grant \#2007CB815406).
   
\clearpage

\begin{deluxetable}{lccccc}
\tablenum{1}
\tablecolumns{6}
\tablecaption{Global Properties of Galaxies Including the Radio 
Continuum in the Gao \& Solomon (GS04a) HCN Survey Sample}

\tablehead{
\colhead{Galaxies}            &       \colhead{$D_L$}          &
\colhead{$L_{\rm IR}$}        &       \colhead{$L_{\rm CO}$}   &
\colhead{$L_{\rm HCN}$\tablenotemark{a}}        &        
\colhead{$L_{\rm RC}$\tablenotemark{b}}  \\    
\colhead{   }                 &       \colhead{Mpc}            &
\colhead{$10^{10}$ \ls}       &  
\multicolumn{2}{c}{$10^8~{\rm K \kms pc}^2$}   &   
\colhead{$10^{5}$ \ls}}         

\startdata

NGC 253 &       2.5 & 2.1& 4.6& 0.27 & 1.53 \\
{\bf IC 1623} &        81.7& 46.7& 130.5& 8.5 & 72.47 \\
NGC 660\tablenotemark{c} &       14.0 & 3.7 &  7.3 & $>$0.26 & 3.31 \\
{\bf NGC 695} &       133.5 & 46.6 & 92.9 & 4.3 & 58.24  \\
{\bf Mrk 1027} &       123.5 & 25.7 & 41.7 & 1.89 & 31.58 \\
NGC 891  &      10.3 & 2.6 & 11.0 & 0.25 &3.25 \\
NGC 1022 &      21.1 & 2.6 &  4.2 & 0.20 & 0.80 \\
NGC 1055  &     14.8 & 2.1 & 13.3 & $<$0.37 & 2.04 \\
{\bf NGC 1068}\tablenotemark{c}  &     16.7 & 28.3 & 20.7 & 3.61 & 60.81  \\
{\bf NGC 1144}\tablenotemark{c}  &     117.3 &  25.1 &  108.9 & 2.67 & 92.93 \\
{\bf NGC 1365}\tablenotemark{c}  &     20.8 & 12.9 & 58.7 & 3.10 & 10.02 \\
IC 342   &        3.7 & 1.4 & 9.5 & 0.47 & 1.35 \\
{\bf NGC 1614}\tablenotemark{c} &      63.2 & 38.6 & 24.5  & 1.25 & 23.92 \\
{\bf *VII Zw 31} &   223.4 & 87.1 & 125.0 & 9.8 & 90.70 \\
{\bf *05189$-$2524}\tablenotemark{c} &    170.3 & 118.1 & 67.0 & 6.2 & 36.49 \\
NGC 2146 &      15.2 & 10.0 & 12.5 &  0.96 & 10.97 \\
NGC 2276 &      35.5 & 6.2 & 10.2 & 0.40 & 15.58 \\
{\bf Arp 55}  &       162.7 & 45.7 & 125.0 & 3.8  & 42.56 \\
NGC 2903 &      6.2 &  0.83 & 2.3 & $>$0.09 & 0.68 \\
{\bf *UGC 05101}\tablenotemark{c} &    160.2 & 89.2 & 50.8 & 10.0 & 190.72 \\
M82    &        3.4 & 4.6 & 5.7  &0.30 & 3.87 \\
NGC 3079\tablenotemark{c}  &     16.2 & 4.3 & 24.0 & $\sim$ 1.0 & 9.73 \\
{\bf *10565+2448} &    173.3 & 93.8 & 61.5 & 10.2 & 74.79 \\
{\bf Arp 148} &       143.3& 36.5& $>$47.0 &4.0& 32.66 \\
NGC 3556 &      10.6& 1.35& $>$4.5& $>$0.09 & 1.50 \\
NGC 3627 &      7.6 &1.26 &4.4 &$>$0.08 & 1.16 \\
NGC 3628 &      7.6 &1.01 &7.1 &0.24 & 1.32 \\
NGC 3893 &      13.9 & 1.15& 4.1& 0.23& 1.17 \\
NGC 4030 &      17.1 &2.14 &15.2 &0.54 &1.97 \\
NGC 4041 &      18.0 &1.70 &3.9 &0.18 &1.45 \\
NGC 4414 &      9.3& 0.81& 4.6 &0.16& 0.91 \\
NGC 4631 &      8.1& 2.0 &2.3 & $\sim$0.08 & 3.44 \\
NGC 4826 &      4.7& 0.26& 1.3& $>$0.04 & 0.11 \\
NGC 5005 &      14.0 &1.4& 8.2& 0.41& 1.51 \\
NGC 5055 &      7.3 &1.1 &8.6& $>$0.10 & 0.91 \\
{\bf NGC 5135}\tablenotemark{c} &      51.7 &13.8 &31.3 &2.73 & 22.65 \\
M83   &         3.7 & 1.4 &8.1 &0.35 & 1.46 \\
{\bf *Mrk 273\tablenotemark{c}}  &      152.2 &129.9& 65.0 &15.2 & 146.44 \\
NGC 5678 &      27.8 &3.0 &17.2& 0.75 & 3.70 \\
NGC 5713 &      24.0& 4.2 &8.1 &0.22& 3.98 \\
NGC 5775 &      21.3& 3.8 &10.9 &0.57 & 5.55 \\
{\bf *17208$-$0014} &    173.1 & 234.5 & 146.9 & 37.6 & 107.08 \\
{\bf 18293$-$3413} &     72.1& 53.7 &85.5& 4.03 &51.35 \\
{\bf NGC 6701} &      56.8 & 11.2 & 34.0 & 1.38 & 13.00 \\
{\bf NGC 6921} &      60.3 &11.4 & 17.5  & $\sim$2.81 & 4.35 \\
NGC 6946 &      5.5 &1.6& 9.2& 0.49 & 1.84 \\
{\bf NGC 7130}\tablenotemark{c} &      65.0 &21.4& 44.9 &3.27 & 33.78 \\
{\bf IC 5179}  &       46.2 &14.1 & 26.4 &3.42 & 15.39 \\
NGC 7331 &      15.0 &3.5& $>$10.7& $>$0.44 & 3.67 \\
{\bf NGC 7469}\tablenotemark{c} &      67.5 &40.7 &37.1 &2.19 & 35.93 \\
NGC 7479\tablenotemark{c} &      35.2 &7.4& 26.7 &1.12 & 5.36 \\
{\bf *23365+3604} &   266.1 &142.0 &85.0& 15.0& 76.10 \\
{\bf Mrk 331}\tablenotemark{c}  &       75.3& 26.9 &52.1 &3.35& 17.51 \\
          &       &      &        &         &           \\
\cutinhead{HCN Data from Literature}
{\bf *Mrk 231}\tablenotemark{c}   &     170.3& 303.5& 82.2 &18.6 & 375.04 \\
{\bf *Arp 220}  &     74.7& 140.2 &78.5 &9.2& 78.98 \\
{\bf NGC 6240\tablenotemark{c}} &      98.1 &61.2 &79.0 &11.0& 164.81 \\
{\bf Arp 193}  &      92.7 &37.3 &39.8& 9.5 & 37.92 \\
{\bf Arp 299}\tablenotemark{c} &       43.0& 62.8 &29.0 &2.1 & 54.70 \\
{\bf NGC 7771} &      60.4 &21.4 &90.8& 6.5 & 19.76 \\
{\bf NGC 828} &       75.4 &22.4& 58.5& 1.3& 26.82 \\
NGC 520 &       31.1& 8.5& 16.3& 0.64& 7.45 \\
NGC 3147\tablenotemark{c} &      39.5 &6.2 &59.0 &0.90 & 6.82 \\
NGC 1530 &      35.4& 4.7 &23.0 &0.49 & 4.42 \\
NGC 4945\tablenotemark{c} &      3.7& 2.6 &5.8 & $\sim$0.27&3.95 \\
M51  &          9.6 &4.2& 19.4 &0.50& 6.12 \\

\enddata 

\tablecomments{This table contains all HCN survey data of
Gao \& Solomon (2004a, GS04a) and includes a dozen galaxies
in the literature (almost entirely from Solomon \etal 1992,
but M51 and NGC~4945 are from Nguyen-Q-Rieu \etal 1992 and
Henkel \etal 1994, respectively). LIRGs with \Lir$>
10^{11}\ls$ are in boldface, and ULIRGs with \Lir$\approxgt
10^{11.9}\ls$ are further marked with an asterisk ($\ast$).
The uncertainties of \Lhcn~ are mainly at a level of up to
$\sim$30\%, whereas for several large sources without
sufficient mapping the uncertainties can be as large as
$\sim$50\% ($\sim$20\% calibration error included; see GS04b
for details). The uncertainties of \Lco\ are of level
$\sim$20\%, mainly dominated by the calibration errors. The
uncertainties of \Lir\ and \Lrc\ luminosities are only of a
few percent.}

\tablenotetext{a}{As in GS04b, the 2$\sigma$ upper limit
($<$) is listed for NGC~1055. The lower limits ($>$) are for
nearby galaxies, where we either only detected HCN in the galaxy
central regions or more extensive mapping is still required beyond
the central pointing.}

\tablenotetext{b}{The \Lrc ~data are mainly obtained from
  the NVSS and literature (e.g., Segalovitz 1977; Condon
  1987; Condon \& Broderic 1988; Condon et al. 1990, 1996;
  Wright \& Otrupcek 1990).}

\tablenotetext{c}{Seyfert AGNs mainly based on the classifications given in the NED, excluding 
those of LINERs and probable Seyferts.}

\end{deluxetable}

\clearpage
\newpage

\begin{figure}
\plotone{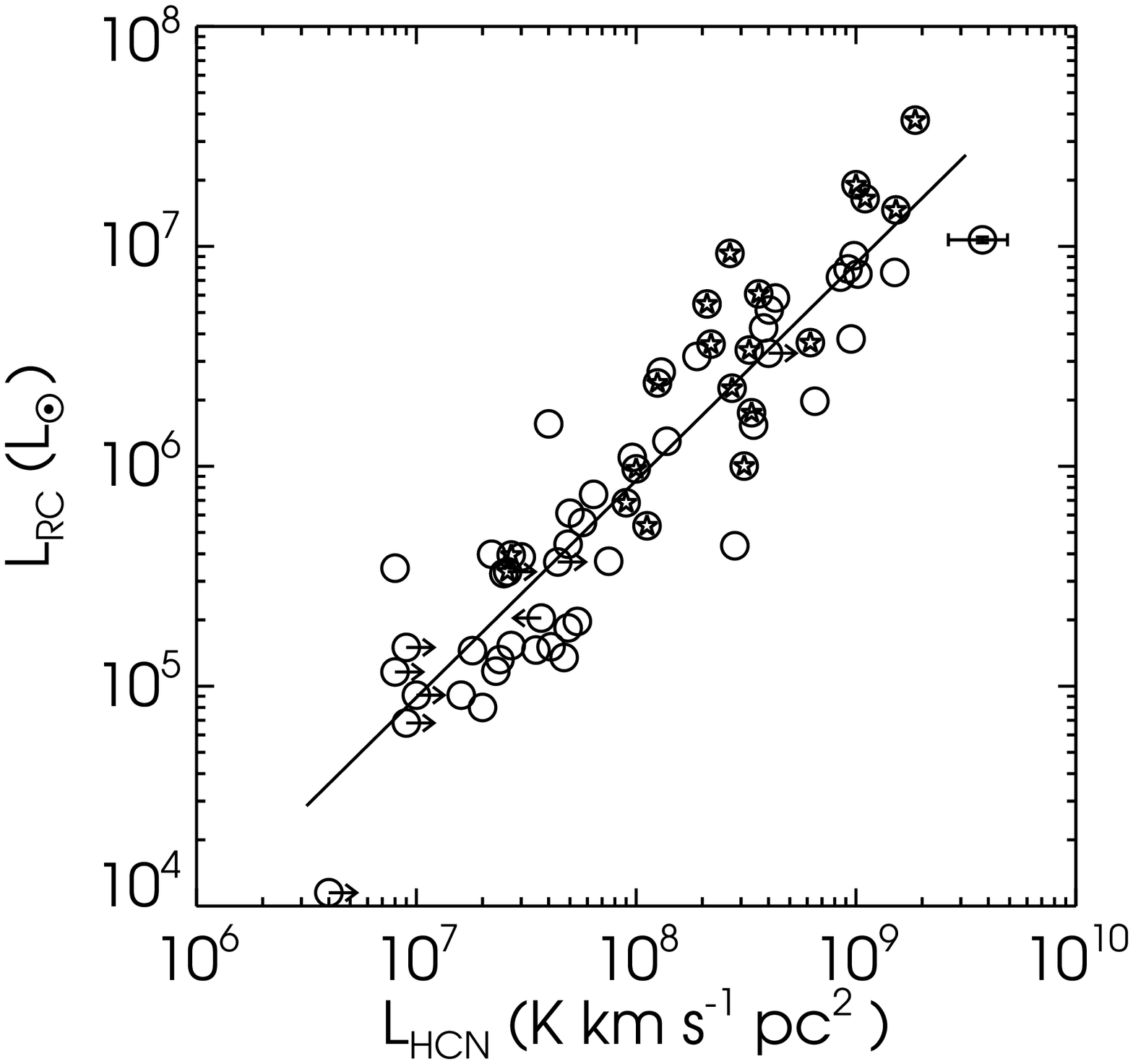}
\end{figure}

\newpage

\begin{figure}
\plotone{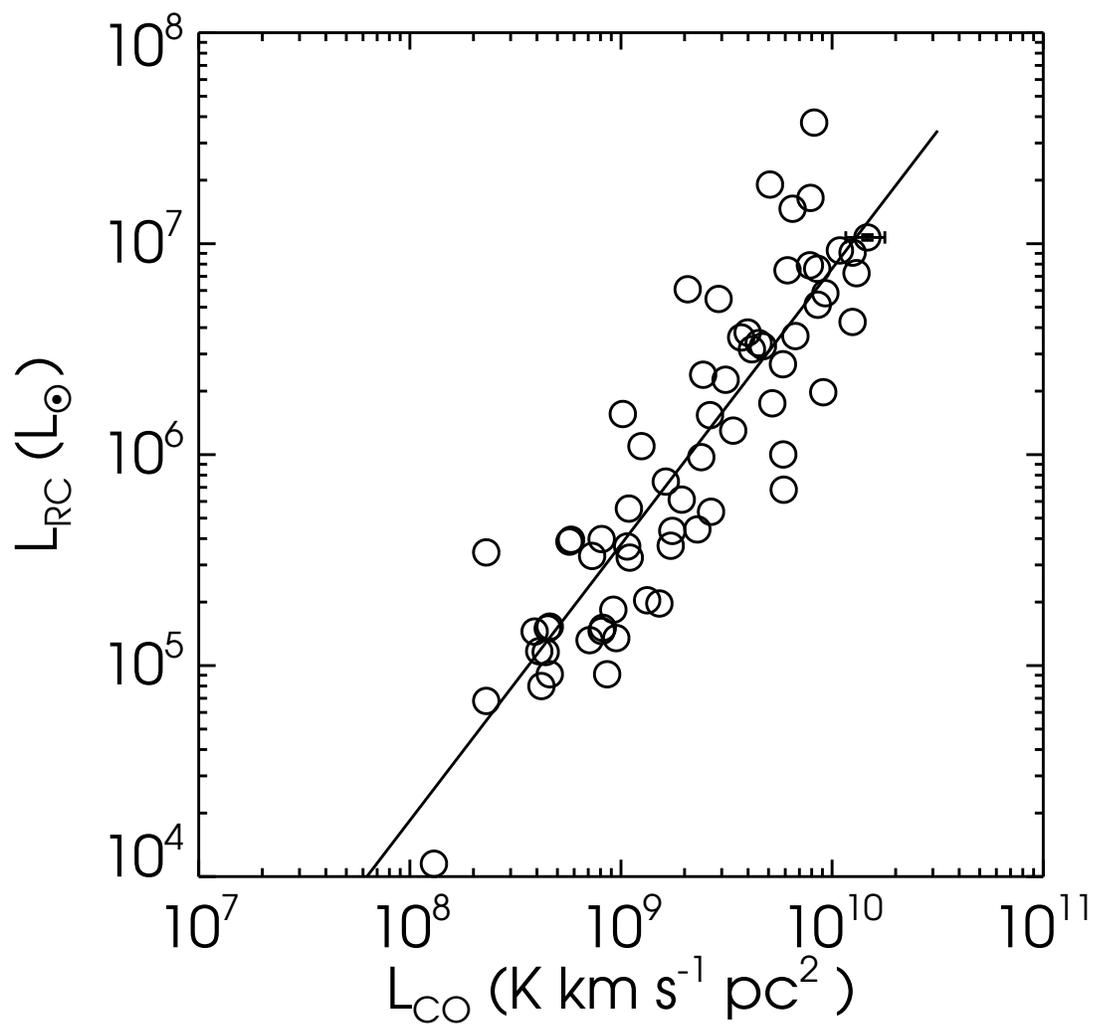} 

\caption{\tiny (a) Correlation between HCN and RC luminosities
in 65 galaxies (the correlation coefficient is $R = 0.92$,
$R^2 = 0.84$). Galaxies with known embedded AGN are
indicated with stars. Some limits in HCN luminosities are
indicated with arrows. The representative error bar
including $~20\%$ calibration uncertainty is shown for the
source IRAS 17208-0014 which has the highest HCN luminosity
in the sample ($\sigma_{\Lhcn} \sim 30\%$, $\sigma_{\Lrc}
\sim 3.1\%$). The fit line: log \Lrc = 0.99 ($\pm $ 0.05) log \Lhcn$ - 2.0$. 
(b) Correlation between CO and RC
luminosities in 65 galaxies. The correlation coefficient is
$R = 0.89$ ($R^2 = 0.79$). The representative error bar
including $~20\%$ calibration uncertainty is also shown for
the source IRAS 17208-0014 ($\sigma_{\Lco} \sim 20\%$). The
fit line: log \Lrc = 1.31 ($\pm $ 0.09) log \Lco$ -6.20$.
\label{fig1}} 
\end{figure}

\newpage

\begin{figure}
\plotone{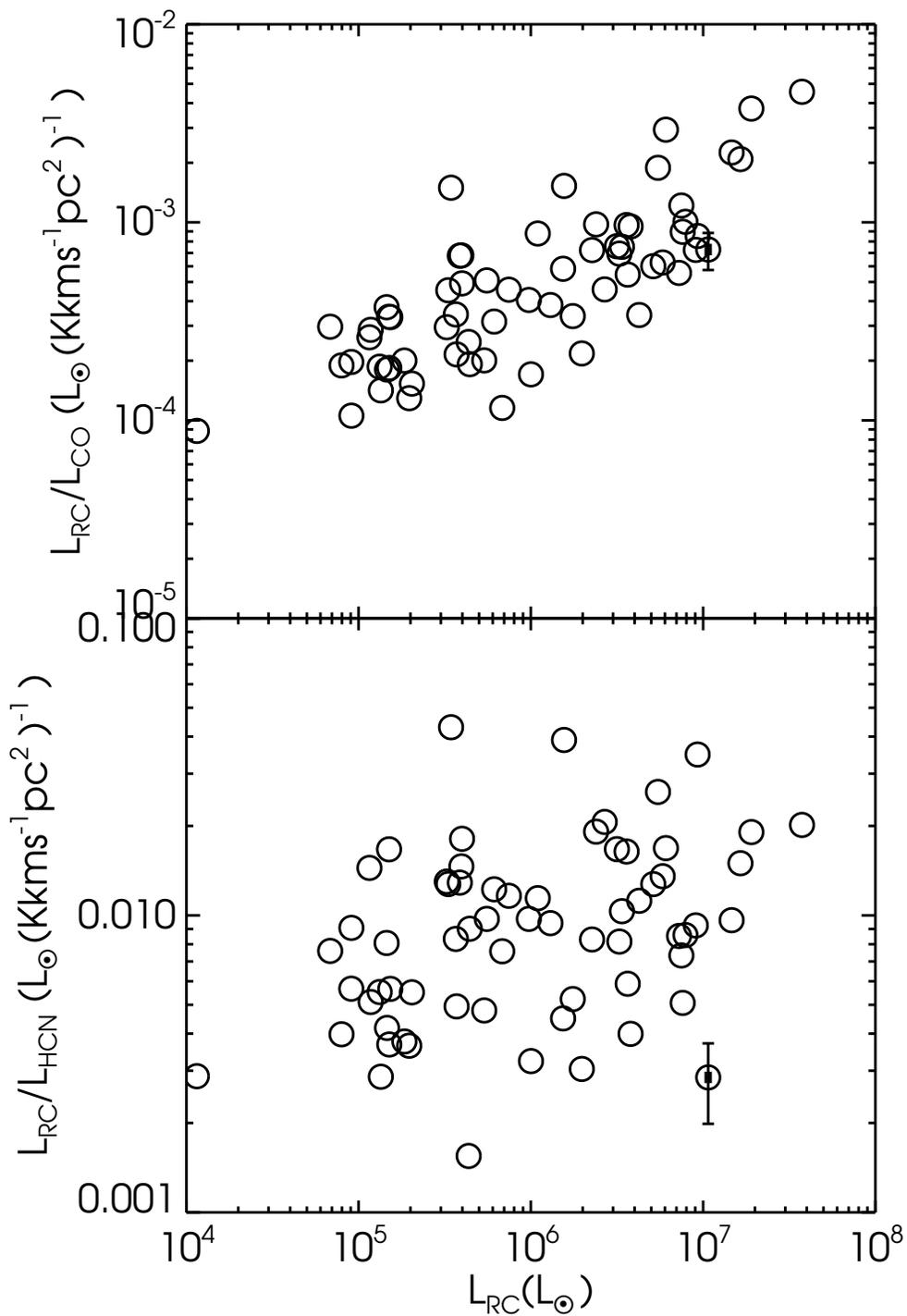} 

\caption{(a) Correlation between RC/CO and RC is
rather prominent, whereas (b) weak or no obvious correlation between
RC/HCN and RC is shown. The representative error bar is
shown for the source IRAS 17208-0014 ($\sigma_{\Lrc / \Lco}
\sim 20\%$, $\sigma_{\Lrc / \Lhcn} \sim 30\%$).
\label{fig2}} 
\end{figure}

\newpage
\begin{figure}
\plotone{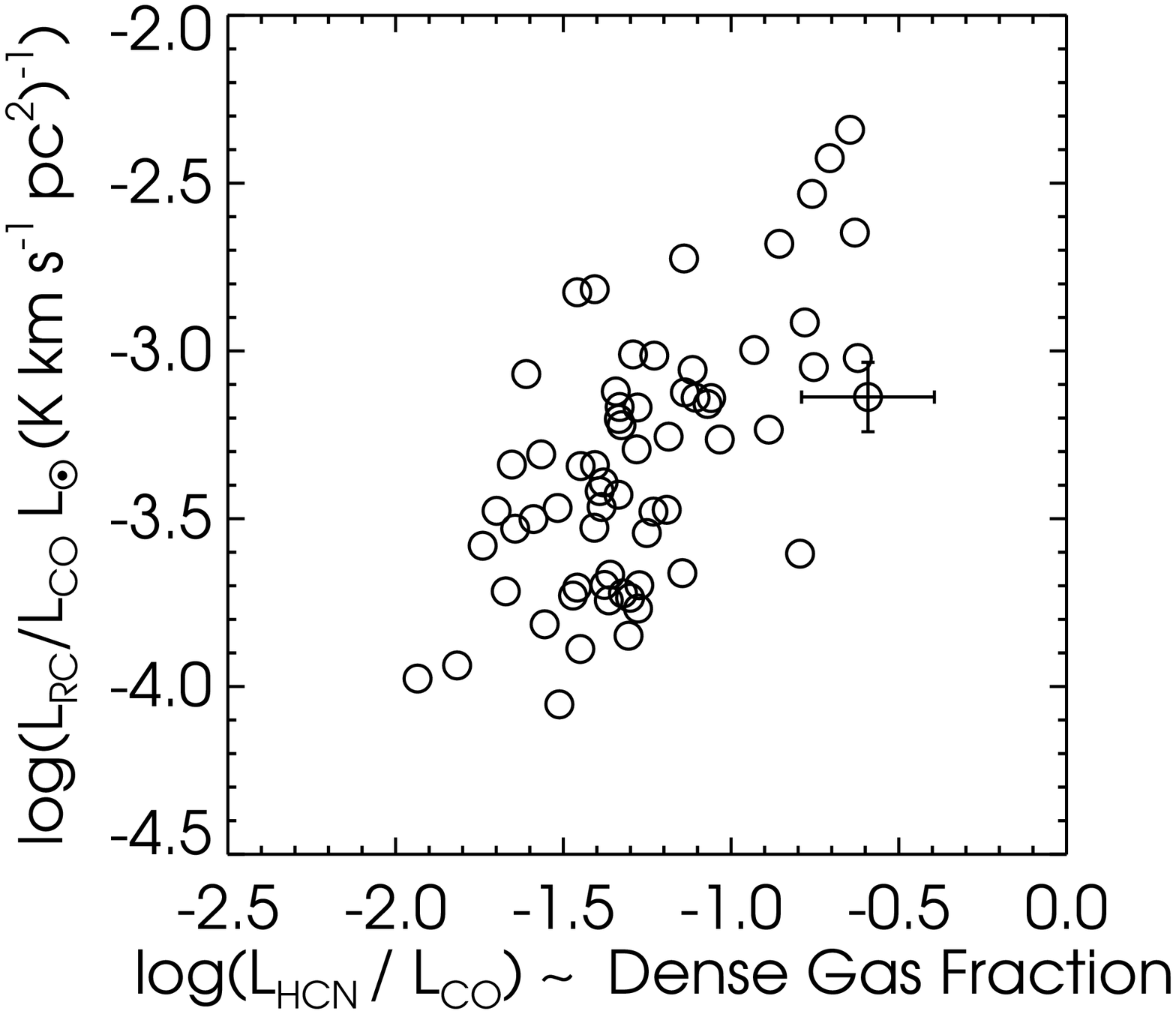}

\end{figure}

\newpage
\begin{figure}
\plotone{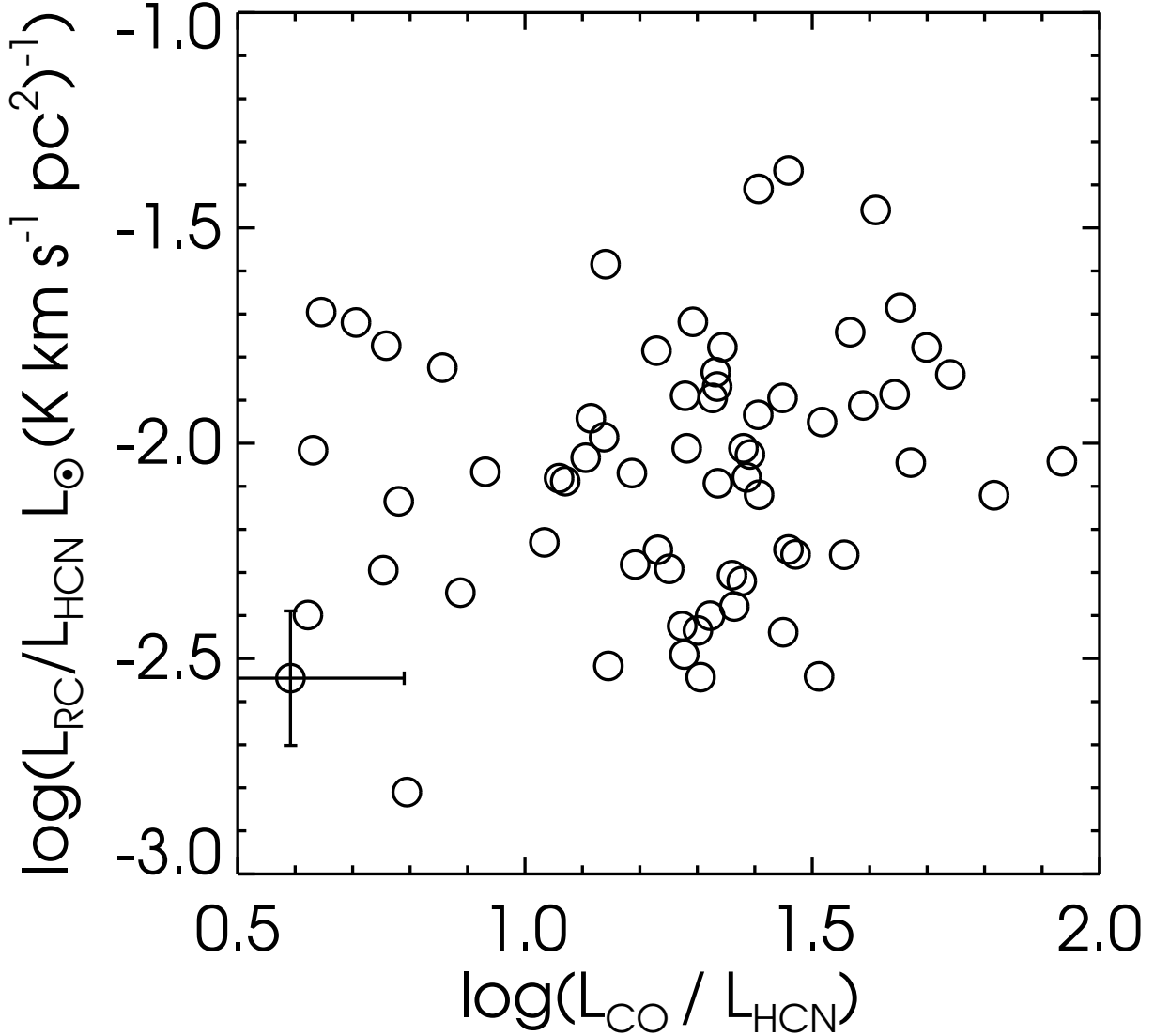} 

\caption{(a) Correlation between RC/CO and HCN/CO shows that
there is indeed a tight correlation between RC and HCN since both
luminosities are normalized by \Lco. The correlation
coefficient is $R = 0.63$ ($R^2 = 0.40$). (b) Normalized by
\Lhcn, the correlation between RC and CO (Figure 1(b)) has
completely disappeared ($R = 0.20$, $R^2$ = 0.04). The
uncertainties of the various ratios are $\sigma_{\Lrc /
\Lco} \sim 21\%$, $\sigma_{\Lrc / \Lhcn} \sim 30\%$, and
$\sigma_{\Lhcn / \Lco} \sim 37\%$.
\label{fig3}} 
\end{figure}

\clearpage

\newpage
\begin{figure}
\plotone{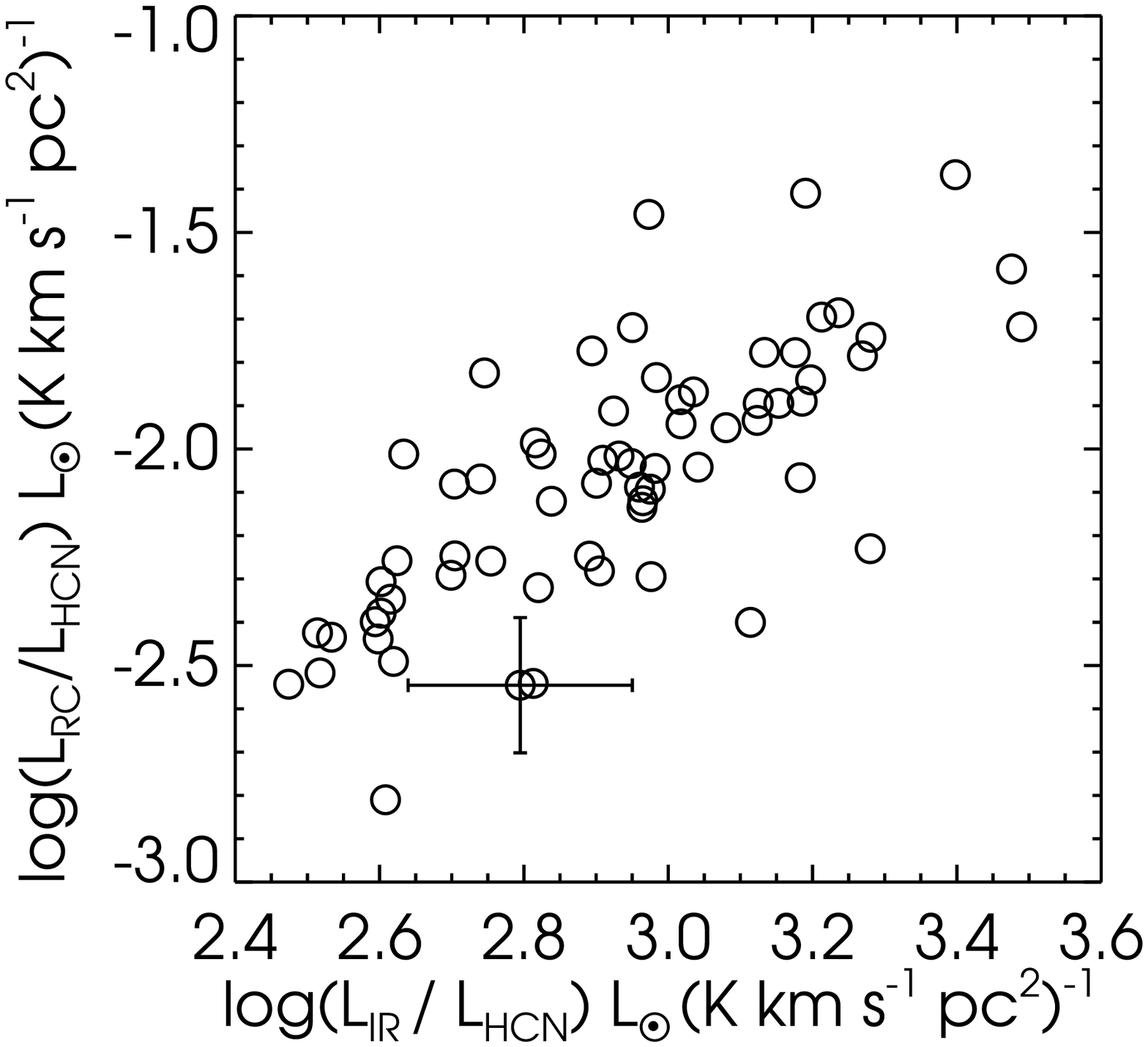}
\end{figure}
\newpage

\begin{figure}
\plotone{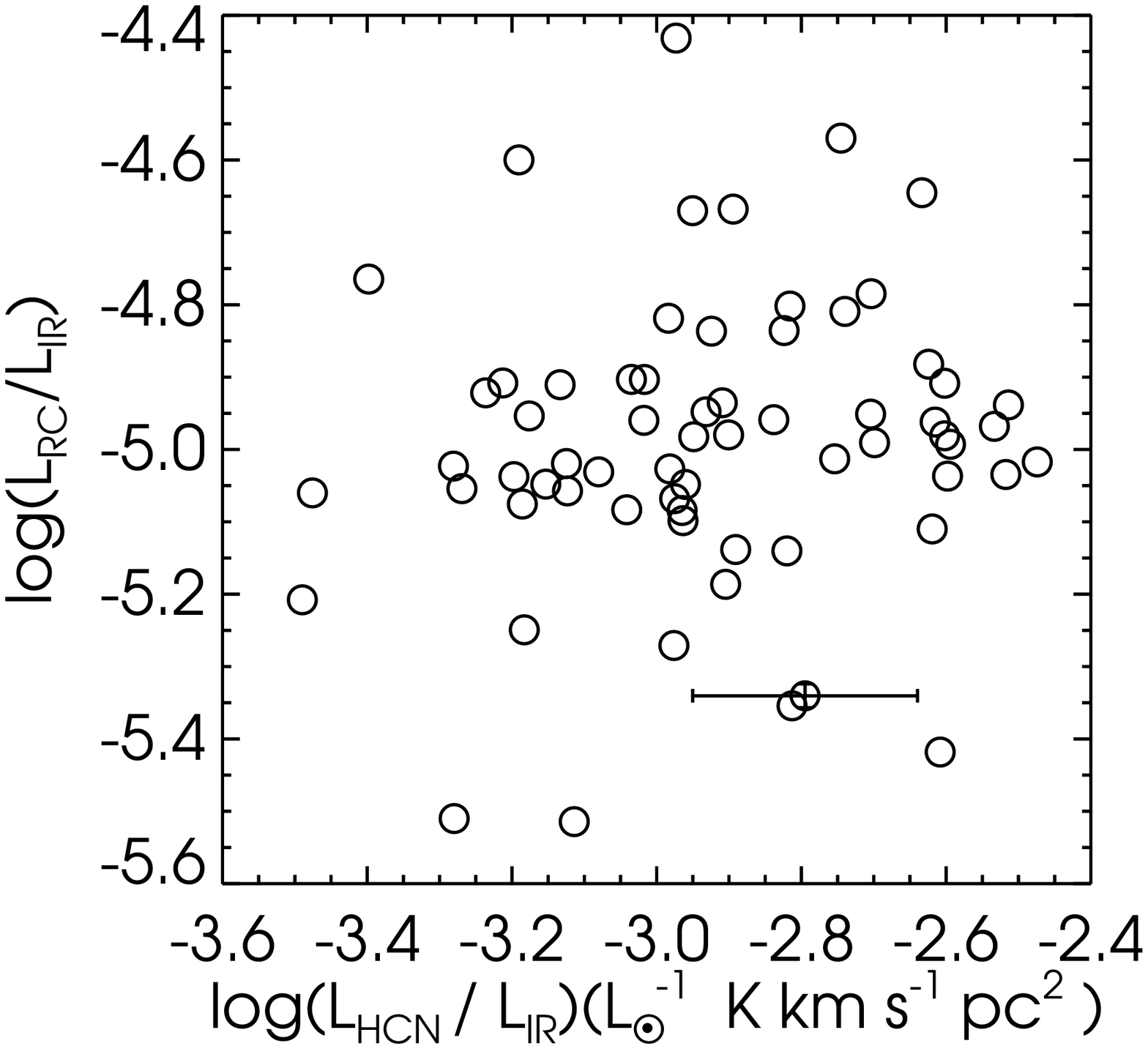}

\end{figure}

\newpage

\begin{figure}
\plotone{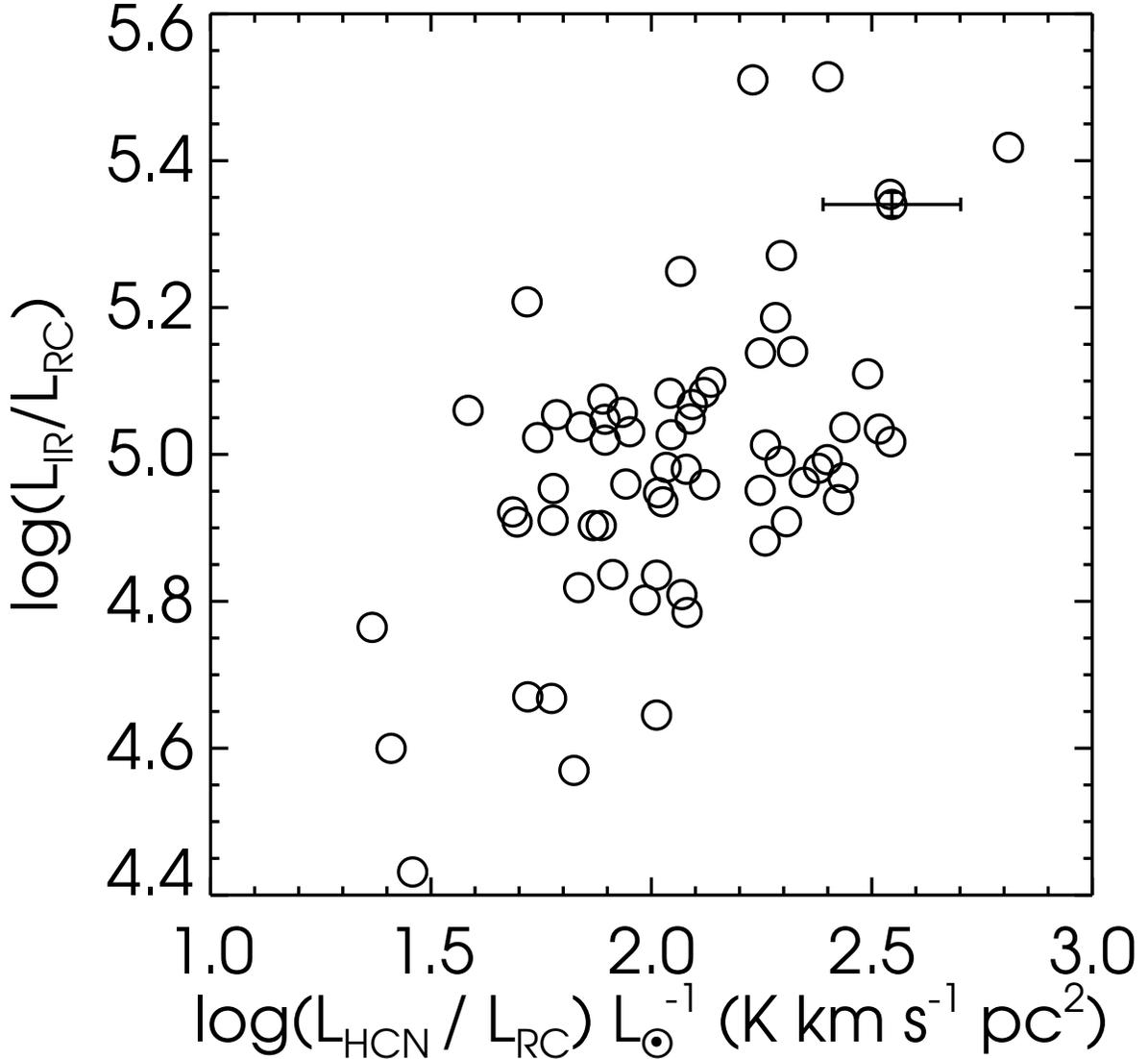}

\caption{(a) Correlation between RC and FIR is very prominent even after
  normalization by \Lhcn \ ($R = 0.74$, $R^{2}$ = 0.55), which implies that the
  correlation between RC and FIR luminosities is really tight.  (b) Normalized
  by \Lir, the correlation between RC and HCN completely disappeared 
  ($R =0.14$, $R^{2}$ = 0.02). (c) Correlation between FIR and HCN is still
  prominent even after normalization by \Lrc \ ($R = 0.57$, $R^{2}$ = 0.33),
  which suggests that their correlation is really tight.  The uncertainties for the
  various ratios are $\sigma_{\Lir / \Lhcn} \sim 30\%$ and $\sigma_{\Lrc /
    \Lir} \sim 4\%$ (duplicated uncertainties can be found in the captions of
  previous figures).}

\label{fig4}
\end{figure}

\newpage
\begin{figure}
\plotone{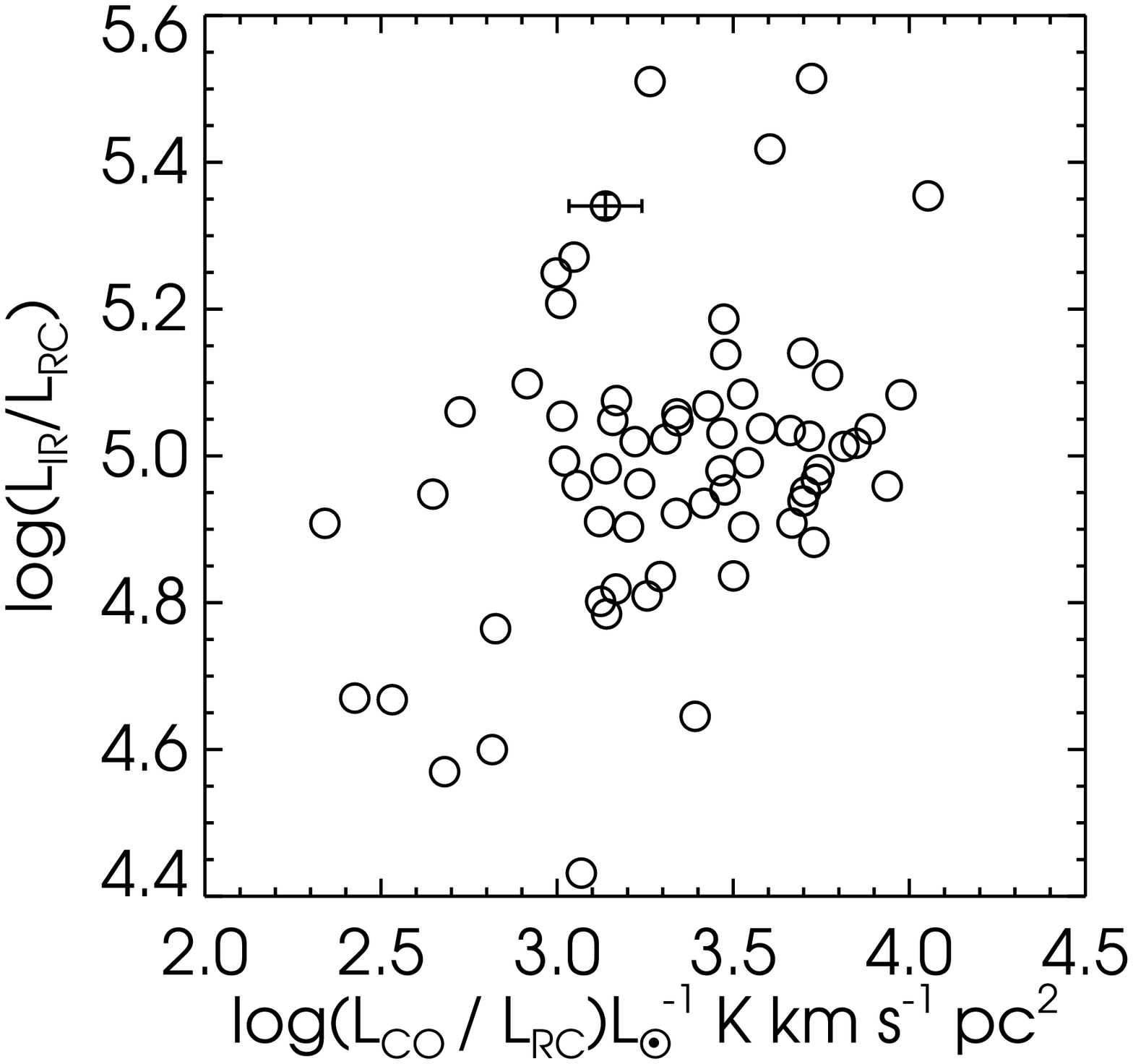}

\end{figure}

\newpage

\begin{figure}
\plotone{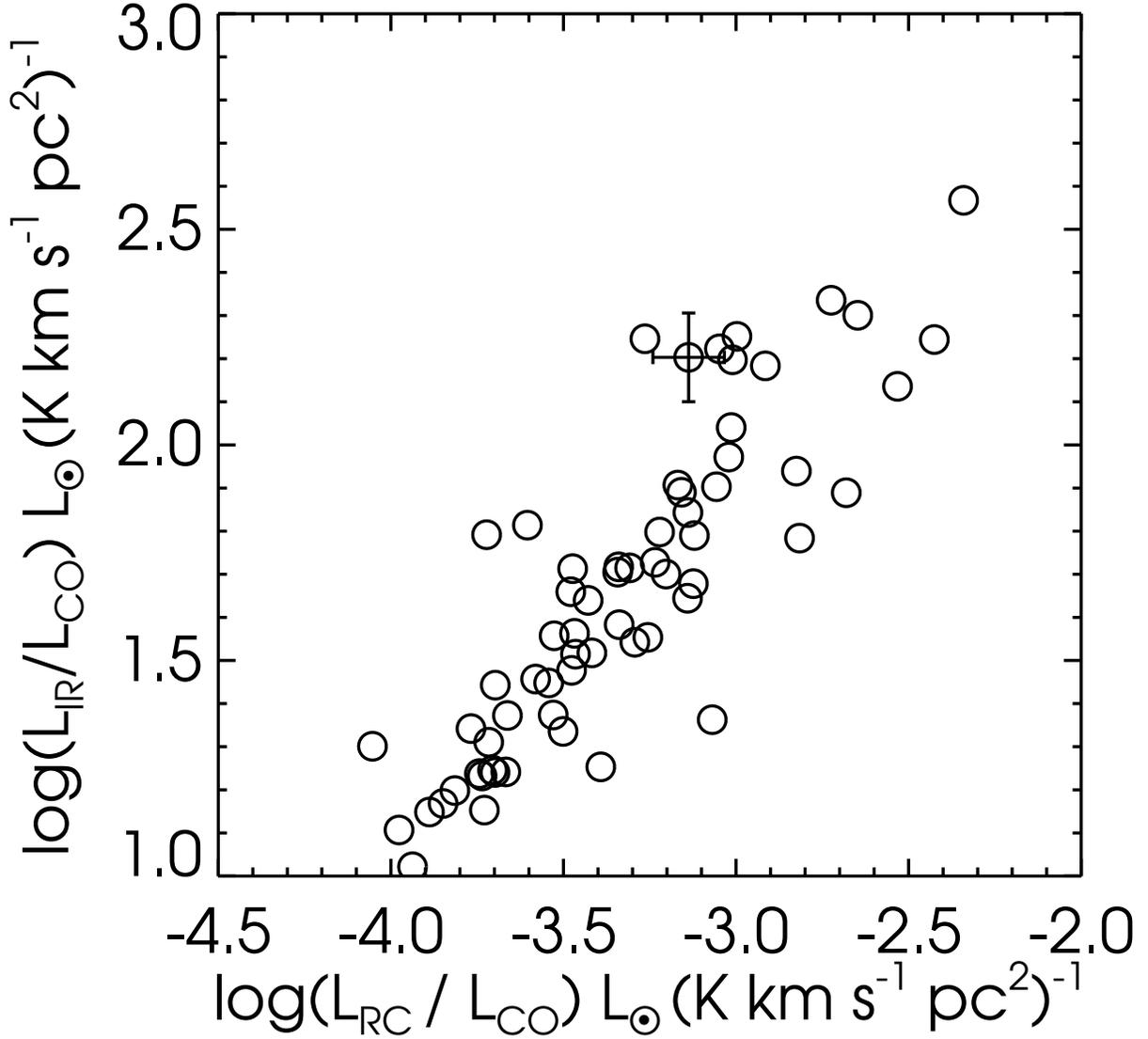} 

\caption{(a) After normalization by \Lrc, the correlation
between FIR and CO is still held, though marginal with $R =0.37$ 
($R^{2}$ = 0.14). (b) Correlation between FIR and RC
is very tight even after normalization by \Lco \ ($R = 0.92$,
$R^{2}$ = 0.85). The uncertainty of $\Lco/\Lrc$ is
$\sigma_{\Lco / \Lrc} \sim 21\%$ (duplicated uncertainties
can be found in the captions of previous figures).
\label{fig5}} 
\end{figure}

\clearpage

\appendix

\section{IR and RC Correlation}

We conclude that among all various correlations, only the FIR-RC and FIR-HCN 
correlations stand out distinguishably better than the rest. Here, we plot 
the well-known FIR-RC correlation in Figure 6 for our HCN sample. Similar to Figure 1(a), 
AGNs are labeled in order to see whether there is any obvious trend in the difference
between AGNs and other galaxies. In particular, we plot other directly related 
correlations such as FIR/RC versus HCN and FIR/RC versus HCN/CO (Figure~7) 
in order to examine any systematic contribution to the possible scatters in the 
tightest FIR-RC correlation.

The obvious trend in Figure 7(a), in which FIR/RC versus HCN emission is plotted, is
that smaller scatter in the FIR/RC ratio (thus a bit tighter correlation) can be 
observed at the lower HCN luminosity end, i.e., the normal galaxies and less 
luminous LIRGs since all ULIRGs have large HCN luminosity. However, for ULIRGs 
and starburst galaxies (i.e., galaxies with HCN/CO$>0.06$, as claimed by GS04a), 
there is an obvious trend of larger scatters in the FIR/RC ratio for higher 
HCN/CO ratio (Figure 7(b)) or high luminosities in HCN or even CO.  The situation is 
quite similar if we show the IR and RC luminosities in the $x$-axis.

In fact, the FIR/RC ratio is not entirely constant and seems to weakly
correlate with the HCN/RC (Figure~4(b)), yet is only marginally dependent on the
CO/RC ratio (Figure~5(a)). Again, this appears to suggest that the extremely large 
scatter in the FIR/RC ratio only exists in galaxies with either the highest or
smallest HCN/RC ratios. 

\section{Other Three-parameter Fits}

In Section 3.3.1, we only listed the three-parameter (FIR, HCN, RC) fits.
The details of the three-parameter (FIR, HCN, CO) fits can be found in GS04a,
which manifest a much tighter linear FIR-HCN correlation versus the 
nonlinear less tight FIR-CO correlation. For the sake of completeness,
we list here two other combinations of three-parameter fits involving 
RC, HCN, and CO and FIR, RC, and CO. First, RC, HCN, and CO:   
\begin{equation}
   log \Lrc(\Lhcn, \Lco) = (0.72 \pm 0.13) log \Lhcn + (0.40 \pm 0.18) log \Lco
   - 3.53.
\end{equation}
The contributing factor from HCN (0.72) to RC is nearly twice as that from CO
(0.40). This is much less than the factor of $>$5 difference between HCN and CO
in the fit to FIR (GS04a).

  $ log \Lhcn(\Lrc, \Lco) = (0.44 \pm 0.08) log \Lrc + (0.69 \pm 0.12) log \Lco - 1.00$, 
   ~and~
\begin{equation}
   log \Lco(\Lrc, \Lhcn) = (0.18 \pm 0.08) log \Lrc + (0.49 \pm 0.09) log \Lhcn
   + 4.28.
\end{equation}
These two fits show that HCN and CO contribute to each other much more than RC does.
Yet, RC's contribution to HCN is quite close to that of CO, whereas RC's contribution 
to CO is significantly less than that of HCN.

And now FIR, RC, and CO:
 $   log \Lrc(\Lir, \Lco) = (0.86 \pm 0.07) log \Lir + (0.21 \pm 0.11) log \Lco + 5.50$, 
 ~and~
\begin{equation}
   log \Lir(\Lrc, \Lco) = (0.79 \pm 0.07) log \Lrc + (0.23 \pm 0.10) log \Lco + 4.09.
\end{equation}
These results are very similar to the results of Equations (2) and (4), demonstrating the tight RC-IR correlation 
with much less contribution from CO. The difference is that the HCN contribution is nearly
comparable to that of RC, much more than that of CO, in the fit to FIR.

\begin{equation}
   log \Lco(\Lir, \Lrc) = (0.33 \pm 0.15) log \Lir + (0.29 \pm 0.14) log \Lrc + 3.92.
\end{equation}
Here, both FIR and RC contribute to CO equally. This is totally unlike Equation (3) where FIR 
contributes most to HCN with much less contribution from RC. Similarly, in Equation (5) 
of GS04a, HCN also contributes much more to CO than FIR.

\newpage
\begin{figure}
\plotone{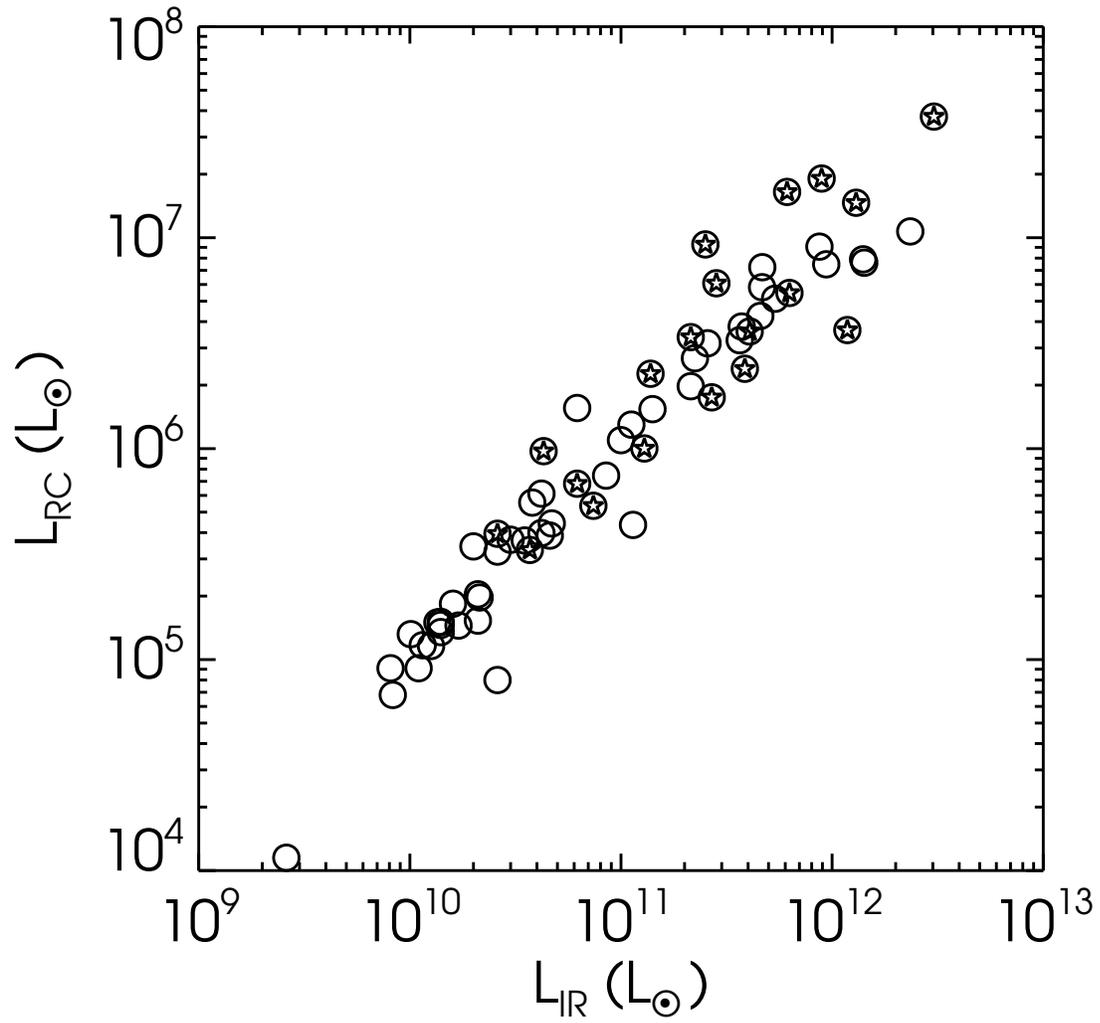}
\caption{Tightest correlation between FIR and RC luminosities in 65 galaxies.
 Galaxies with known AGN embedded are indicated with stars.
\label{fig6}}
\end{figure}

\newpage

\begin{figure}
\plotone{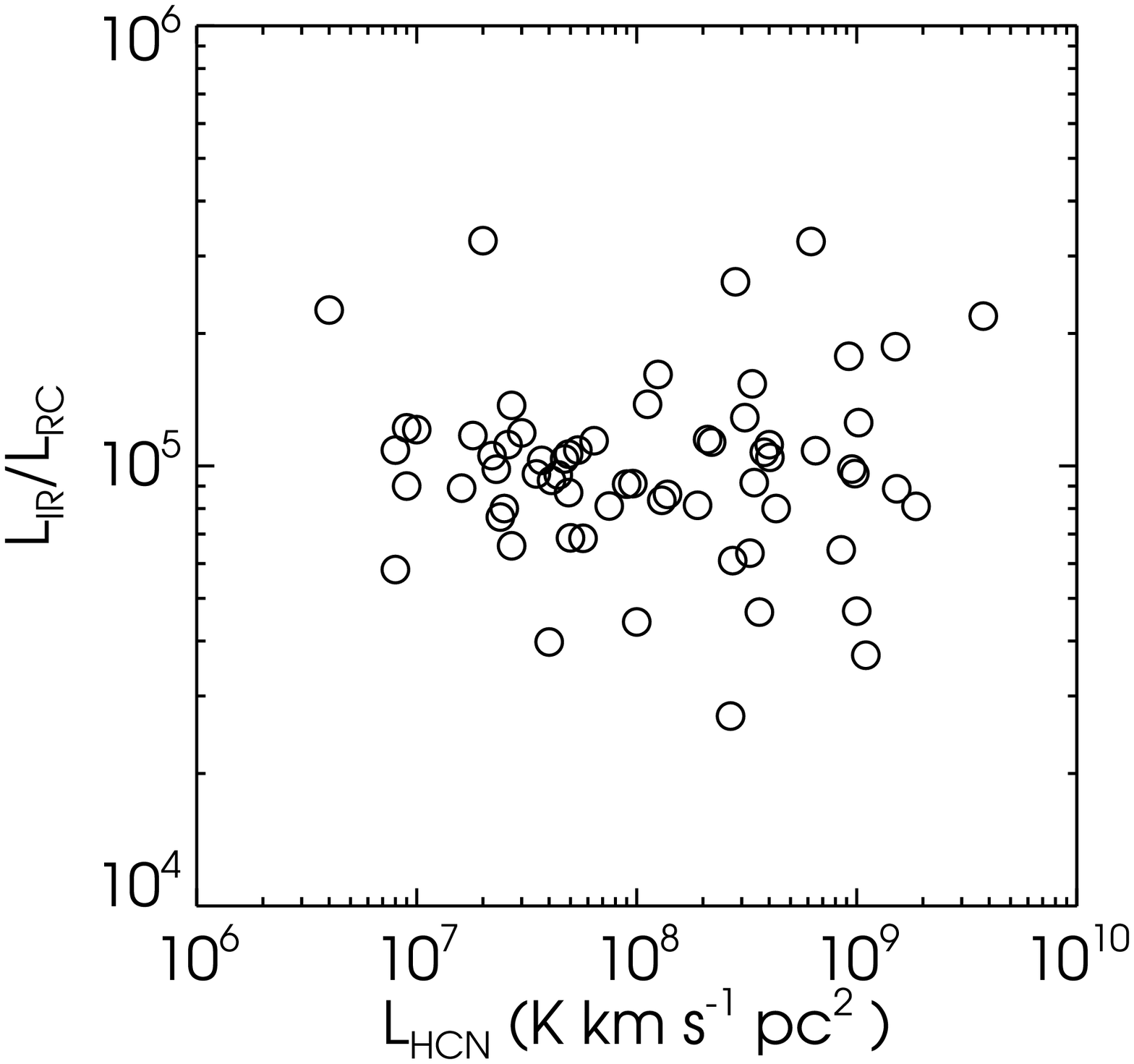}
\end{figure}
\newpage

\begin{figure}
\plotone{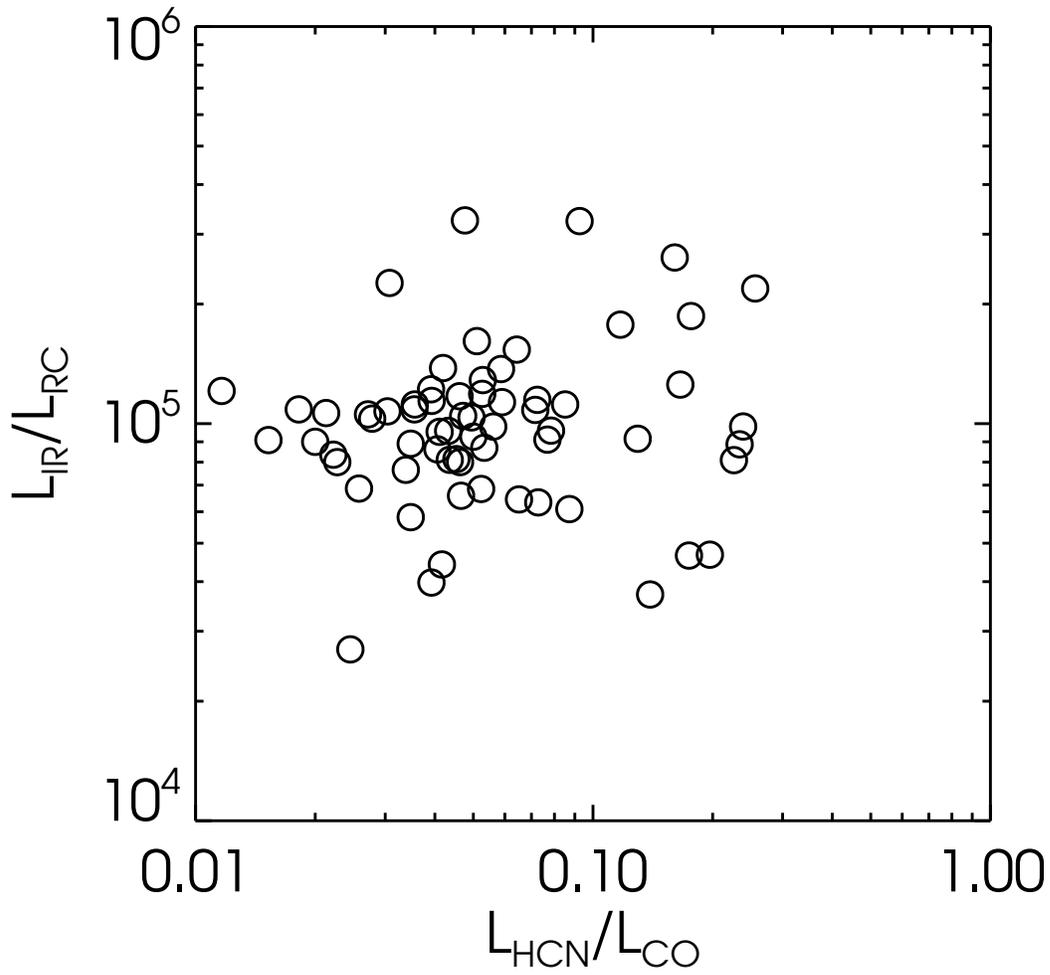}
\caption{(a) No correlations between FIR/RC luminosity ratio and HCN line
 luminosity, and (b) between the  FIR/RC and  the HCN/CO luminosity ratio. However,
 larger scatter in the FIR/RC luminosity ratio appears to exist in galaxies with high 
 HCN luminosity and larger HCN/CO ratios.
\label{fig7}}
\end{figure}

\end{document}